\documentclass[phd, titlesmallcaps, examinerscopy, copyrightpage, foronline]{mqthesis}

\usepackage{lscape}
\usepackage{rotating}  
\usepackage{theorem}   
\usepackage{bm}  
\usepackage{amsmath,amsfonts,amssymb}
\usepackage{booktabs}  
\usepackage{pdfsync}   
\usepackage{graphicx}
\usepackage{latexsym}
\usepackage{amsmath}
\usepackage{amssymb}
\usepackage{fancyvrb}
\usepackage{subfigure}
\usepackage{colortbl}
\usepackage{enumerate}
\usepackage{pifont}
\usepackage{stmaryrd}
\usepackage{textcomp}
\usepackage{fncylab}
\usepackage{multirow}
\usepackage{paralist}
\usepackage{wrapfig}
\usepackage{colortbl}
\usepackage{longtable}
\usepackage[]{algorithm2e}
\usepackage[table]{xcolor}
\usepackage[protrusion=true,expansion=true]{microtype}
\usepackage[T1]{fontenc}
\usepackage{lscape}
\usepackage{lipsum}

\begin{document}

\frontmatter

\title{Social Influence and Radicalization:\\A Social Data Analytics Study}

\ifthenelse{\boolean{foronline}}{
  \author{\href{mailto:Vahid.moraveji-hashemi@hdr.mq.edu.au}{Vahid Moraveji Hashemi}}
  \department{Computing}
}{
  \author{Vahid Moraveji Hashemi}
  \department{Computing}
}
\degrees{Master of research}

 \submitdate{October 2018}

\renewcommand{\degreetext}%
{Master of research}

\titlepage

\chapter{Dedication}

To my lovely parents and my elder brother Alireza for their love, patience, and support. 

\chapter{Acknowledgements}

Working at the department of Computing at the University of Macquarie has been a great pleasure and a wonderful privilege.

In the first place, I would like to express my sincere appreciation and deep gratitude to my supervisor, Dr. Amin Beheshti, for his exceptional support, encouragement and guidance during the last year. Amin taught me how to do high quality research and helped me think creatively. His truly incredible academic excellence and beautiful mind have made him as a constant oasis of novel ideas and passions in science, which has inspired and enriched my growth as a student and a researcher. 
Moreover, I thank him for providing me with the opportunity to work with a talented team of researchers.

I am thankful to everyone in the department of Computing at Macquarie University, Data Analytics Research Group, and my friends Adnan Mahmood and Zawar Hussain for their 
help and support during my Master by Research studies. 

I am deeply and forever indebted to my parents and my elder brother Alireza for their love, patience and support. \\
\\
\\
Vahid Moraveji Hashemi\\
Sydney\\
Australia\\
October 2018

\chapter{DISSERTATION EXAMINERS}

\begin{itemize}
\item[$\bullet$] Prof. Salil Kanhere, UNSW Sydney, Australia

\item[$\bullet$] Prof. Aditya Ghose, University of Wollongong, Australia
\end{itemize}

\chapter{Abstract}

The confluence of technological and societal advances is changing the nature of global terrorism. For example, engagement with Web, social media, and smart devices has the potential to affect the mental behavior of the individuals and influence extremist and criminal behaviors such as Radicalization. In this context, social data analytics (i.e., the discovery, interpretation, and communication of meaningful patterns in social data) and influence maximization (i.e., the problem of finding a small subset of nodes in a social network which can maximize the propagation of influence) has the potential to become a vital asset to explore the factors involved in influencing people to participate in extremist activities.

To address this challenge, we study and analyze the recent work done in influence maximization
and social data analytics from effectiveness, efficiency and scalability viewpoints. We introduce
a social data analytics pipeline, namely iRadical,
to enable analysts engage with social data to
explore the potential for online radicalization. In iRadical, we present algorithms to analyse the social data as well as the user activity patterns to learn how influence flows in social networks.
We implement iRadical as an extensible architecture that is publicly available on GitHub and present the evaluation results.

\tableofcontents
\listoffigures
\listoftables

\mainmatter

\chapter{Introduction}
\label{chap:introduction}

Communication has always been a desire for humans. Thus, it has resulted in many inventions. In the communication process and sharing ideas and knowledge, a person may have an influence (i.e., an impact on the character, development, or behavior) on others. This social influence has drawn the attention of different groups in the area of psychology, marketing and politics. The aim of this thesis is to survey social influence and one of its applications called radicalization as well as studying social data analytics with the goal of understanding the context and activities of each user in a social network.

Social influence has been defined as an alteration in one's ideas, feelings, views, and behaviors that extracted from communication with another person~\cite{rashotte2007social}.
A key challenge in 
this area
is the problem of \emph{influence maximization}. This challenge is basically inspired by viral marketing applications, and 
first defined by Kempe et al.~\cite{kempe2003maximizing} as follows:
given a social network which can be shown by a directed graph\footnote{
A graph in a social network consists of nodes and edges, where nodes refer to users and edges refer to social ties between users. A directed graph is a graph whose edges have only one direction (i.e., given a directed graph G, if node A $\rightarrow$ node B, it means that node A has a link to node B, but node B does not have any link to node A).},
where nodes represent users, edges represent social links and influence probabilities are represented by edge weights.
Here, the objective is to choose a group of $k$ users, so that the expected influence spread (which can be represented by expected number of influenced users) is maximized. The expected influence spread of seed users is dependant on the influence diffusion process. This process is captured by a propagation model or a model for influence propagation. In other words, how influence propagates or diffuses across the network, is governed by the propagation model.

Besides viral marketing, social influence can be used in many social network analytics applications such as radicalization, recommender systems, fake news and more. In this dissertation, our focus is on radicalization. While radical thinking itself is not problematic, it turns into a threat to security of a nation when it leads to violence. It would be very challenging to understand the patterns of extremist and criminal activities on social networks, nonetheless, there are some general elements in knowledge of most individuals who have become radicalized, notwithstanding their opinions and incentives~\cite{radical}. Radicalization~\cite{kruglanski2014psychology} - the procedure by which a group, or an individual comes to adopt gradually severe social, political, or religious aspirations and ideals - contains two main dimensions:
(i)~Personality: the motivational and ideological module that describes a purpose to which one may be obligated and recognizes the means of regressive behavior; and
(ii)~Influence: the social procedure of group dynamics and networking, where cyber-enabled propaganda campaigns are increasingly used to exert strategic influence.

Major barriers to the effective understanding of extremist and criminal behaviors on social networks includes the ability to understand the context as well as eventual influence that an influence campaign may have, given an understanding of the mode of information dissemination and indicators that identify the use of artificial promotion of a theme.
This task is challenging and requires technique for:
(i)~the discovery, interpretation, and communication of meaningful patterns in social data (i.e. social data analytics) to explore the potential for understanding and analyzing the personality dimension; and
(ii)~finding a few number of nodes in a social network that can increase the propagation of influence, i.e., Influence Maximization~\cite{chen2009efficient}, for understanding and analyzing the influence dimension.

In this dissertation, we analyze the state-of-the-art related works in influence maximization from three dimensions: Effectiveness, Efficiency and Scalability. We also analyze the state-of-the-art in social network analytics, as it plays a vital role in understanding the context and activities of users in a social network. The content generated on social networks can be analyzed by different methods of machine learning and natural language processing so that we can find out how significant that context would be in affecting other users. In addition to the context, the activities of users
(e.g., number of followers, friends, posts, likes and comments) 
can be assessed as well to help analysts realize which users are more influential in a social network.
From mathematical viewpoint, there are several mathematical diffusion models regarding propagation of information, two of which are important~\cite{chen2013information}:
Independent Cascade Model and Linear Threshold Model.
These models help analysts understand and model information diffusion mathematically.

%

\section{Key Research Issues and Contributions Overview}

In this section, we outline key research issues tackled in this dissertation.
We intend to enabling the analysis of influence maximization and context analytics in online social networks with the goal of identifying extremist behaviors.
We therefore separate research issues into two areas:
(i)~Influence Maximization: analyze the user activity patterns to learn how influence flows in social networks; and
(ii)~Social Network Analytics: analyze posts and comments generated by social actors, from sociological and psychological perspectives

To address these research issues, we study and analyze the recent work done in influence maximization and social data analytics from effectiveness, efficiency and scalability viewpoints.
We present a social data analytics pipeline, namely iRadical, to allow analysts to get involved in social data to explore the potential for online radicalization.
In iRadical, we present algorithms to analyze:
(i)~the user activity patterns to learn how influence flows in social networks; and
(ii)~social data, e.g., posts and comments generated by social actors, from sociological and psychological perspectives.
We introduce the novel concept of Social Personality Graph, to model and analyze factors that are driving extremist and criminal behavior.
The Personality Graph is time-aware and enable analysts to understand and predict the eventual influence of a potential influencer as well as an influence campaign.
We implement iRadical as an extensible architecture that is publicly available on GitHub\footnote{https://github.com/DataAnalyticsResearchGroup/iRadical}.
We present the evaluation results on the performance and the quality of the results using real-world and synthetic data.

The remainder of this dissertation is organized as follows.
We start with a discussion of the current state of the art in influence maximization and social network analytics in ~\autoref{chap:LR}.
%
After that, we present the details of our framework for exploring the potential for online radicalization in ~\autoref{chap:Proposed Model}.
In this chapter, we present our optimized algorithm for influence maximization in social networks as well as a data analytics pipeline to allow analysts to explore the potential for online radicalization.
%
In ~\autoref{chap:experiments}, we provide the implementation and discussion about the evaluation results on the performance and the quality of the results using real-world and synthetic data.
Finally, in~\autoref{chap:conclusion}, we give concluding remarks of this dissertation and discuss
possible directions for future work.

\chapter{Background and State-of-the-Art}
\label{chap:LR}

The confluence of technological and societal advances is changing the nature of global terrorism. For example, engagement with Web, social media, and smart devices has the potential to affect the mental behavior of the individuals and influence on extremist and criminal behaviors such as Radicalization.
In this context, social data analytics (i.e., the discovery, interpretation, and communication of meaningful patterns in social data) and
influence maximization (i.e., the problem of finding a small subset of nodes in a social network that could maximize the spread of influence)
has the potential to become a vital asset to explore the factors involved in influencing people to participate in violent extremism activities.
This chapter therefore presents central concepts and the current state-of-the-art in influence maximization and social data analytics.

\section{Influence Maximization}
\label{Detection of cyber-enabled influence maximization}

Influence maximization, i.e., the problem of discovering a small subset of nodes in a social network which can maximize the
propagation of influence~\cite{chen2009efficient}, has become an effective method for bringing disconnected
offline users together in social networks and connecting people.
In this section, we discuss the related work in influence maximization and classify the existing approaches based on 
effectiveness (reaching a high quality result), efficiency (performing algorithms quickly) and scalability (performing without changes in the quality of result for different size of dataset) of the proposed algorithms.

\subsection{Influence Maximization: Effectiveness }
\label{Effectiveness}

A set of algorithms 
have been 
presented to improve the effectiveness of influence maximization.
Liu et al.~\cite{liu2012time} developed an algorithm in terms of time restriction for the impact maximization problem. In their prototype, they demonstrated that the issue is NP-hard and they also showed that the issue of time limitation influence maximization is monotonic and submodular. By their experiments on four datasets they showed the effectiveness of their algorithm compared to greedy algorithm.

Zhuang et al.~\cite{zhuang2013influence} investigated the problem of impact maximization in a non-static social network. They developed a novel algorithm to estimate the best solution. Their algorithm minimized the mistakes among perceived network and real network. They also applied their algorithm on dissimilar real datasets and their results reveal that the efficiency of their algorithm is comparatively better than other proposed algorithms.

Song et al.~\cite{song2015influence} proposed an algorithm for extracting nodes which have significant impact. Their algorithm includes two sections: separating social network which is large-scale in some groups through considering information spread and choosing groups to identify dominant nodes. Afterwards, for enhancing the performance, they 
performed the influence spread in terms of communities in a paralleled way and examined the influence spread in groups. They also analyzed the estimation guarantees of their models. Their experimental results on real social networks reveal that their developed algorithms can outperform based on performance and accuracy compared with previous methods.

Most of current literature studied the maximization of influence problem only in terms of a distinct network. Nonetheless, these days users often connect to several social networks such that information could be propagated through different networks at the same time. Thus, Zhang et al.\cite{zhang2016least} proposed an uniform framework to analyze and describe the influence spread in multiple networks. They tried to solve the maximization of influence problem by plotting a number of networks into a single one through strategies with and without loss coupling. The strategy without loss coupling keeps all features of primary networks to obtain high-standard solutions while the strategy with loss coupling gives a fascinating option when the execution time and memory usage are main issues. They 
performed 
a set of 
experiments on various datasets, which showed the effectiveness of their framework.

Gong et al.~\cite{7515281} developed a memetic algorithm for group-based maximization of influence to enhance the 2-skip influence propagation to identify the most influential nodes. They particularly used two algorithms, initialization of population and local search, to speed up the convergency of the algorithm. 
In most studies on maximization of influence, the problem that who is affected and how varying the affected population is, has greatly been overlooked. However, Tang et al.~\cite{tang2014diversified} studied the importance of influence and varying the affected population at the same time. First, they presented a framework for influence maximization by creating a class of variety to measure the variety of affected population. Afterwards, they proved that a greedy algorithm can obtain an excellent solution. Furthermore, they concentrated on variety of nodes selected for beginning activation and they also demonstrated how this approach can work for many heuristic algorithms. Finally their experiments on various datasets illustrated that their framework is successful in producing variant results.
Gong et al.~\cite{gong2016influence} presented a particle swarm optimization algorithm(PSO) to discover the most dominant nodes in a social network. The rules for particles in PSO algorithm were restated in their developed algorithm. Their experimentations on four different datasets showed the efficacy and performance of their algorithm. 

Group detection through maximization of influence in social networks is still an issue which has been taken into account in many research studies. Jiang et al.~\cite{jiang2014uniform} 
presented a framework in terms of maximization of influence to discover groups with both overlapping and arranged in order of rank structures. In their framework, they selected nodes with high local impact to run the experimentations of impact proliferation on real datasets. Their results demonstrated that their framework can discover groups efficiently and effectively.

\textbf{Discussion}
There are some related works which have tried to address the influence maximization problem to discover the most influential users accurately. Most related works highlighted that influence maximization is NP-hard and they have also focused on heuristic algorithms such as
genetic~\cite{townsend2003genetic}, memetic~\cite{gong2016efficient} and PSO (Particle Swarm Optimization)~\cite{marini2015particle}
to solve the influence maximization problem. However, each heuristic algorithm is suitable for different kinds of dataset such as big or small datasets. Therefore, it is possible to choose the best heuristic algorithms by try and error on different datasets to ensure which algorithm suits a specific kind of dataset. In addition, some studies have used different approaches like clustering and classification complementary to heuristic algorithms. The experiments show that merging clustering and/or classification techniques with heuristic algorithms will result in more accurate outcomes.

\subsection{Influence Maximization: Efficiency }
\label{Efficiency}

There are some methods which have been developed to enhance the productivity of impact maximization.
%
%
Kempe et al.~\cite{kempe2003maximizing} demonstrated that influence maximization using Linear Threshold model is NP-hard and they also showed that a natural greedy algorithm can acquire optimal result (about 63\%) for different models. This algorithm has still some drawbacks regarding performance.
Therefore, Goyal et al.~\cite{goyal2011simpath} offered an algorithm named SIMPATH to solve these shortcomings by leveraging some optimizations. By testing their algorithm on four real datasets they showed that SIMPATH performs better in different criteria such as execution time, memory usage and standard of seed.

Chen et al.~\cite{chen2009efficient} investigated influence maximization from two perspectives. First, improving the greedy algorithm proposed in~\cite{kempe2003maximizing} and~\cite{leskovec2007cost} 
to decrease the running time. Second, they proposed new heuristic algorithm to enhance influence spread. Their investigational results show that their enhanced greedy algorithm has a comparatively good execution time compared to the previous algorithms. Besides, their new heuristic algorithm obtained much better influence spread in comparison with classic heuristics.
Kimura et al.~\cite{kimura2006tractable} proposed two cases of Independent Cascade Models (ICM) to efficiently compute the anticipated number of nodes in the impact maximization problem. Furthermore, they found out if there are a few numbers of link between nodes, ICM can find influential users with better approximation.
Leskovec et al.~\cite{leskovec2007cost} developed an algorithm named CELF for solving the rudimentary greedy algorithm whose number of nodes is quadratic. Goyal et al.~\cite{goyal2011celf++} developed a novel version of CELF called CELF++ which outperforms CELF in terms of efficiency (35-55\% faster).

Chen et al.~\cite{chen2011influence} proposed a supplement to the independent cascade model to merge the disclosure and spread of pessimistic beliefs. Their model has a factor called factor of quality to model pessimistic treatment of individuals saying negative about a product owing to product shortcomings. Their model actually merges negativity bias which are verified in social psychology literature. They also defined a ratio for influence graph which signifies that selection of seed is dependent on the standard factor for general graphs. Finally, they designed a novel algorithm which is efficient. Through experiments, they showed that their algorithm outperforms a standard greedy algorithm with faster orders of magnitude.

Borgs et al.~\cite{borgs2014maximizing} developed an algorithm for the impact maximization problem in the process of the independent cascade model. 
Jiang et al.~\cite{jiang2011simulated} 
developed a novel method in terms of Simulated Annealing 
(i.e., a technique for finding global optimum in any procedure) for the impact maximization problem. 
They presented two heuristic algorithms to speed up the confluence procedure of SA. Through their results in four real-world datasets. 
Chen et al.~\cite{chen2014cim} presented a new approach named Community-oriented Influence Maximization (CIM) to solve the problem of impact maximization. Their approach focuses on time efficiency and the proposed framework includes three 
steps: i)~detection of group; ii)~creating nominee; and iii)~choosing seed.
The first step detects the group structure of the network, while the second step checks the information of each group to minimize the number of groups. The final step, chooses the seed nodes from the nominees. Through their experiments they show that their framework can choose the number of seeds efficiently to boost information propagation. They also demonstrate that CIM can perform better based on efficiency and scalability in comparison with previous methods.

Zhou et al.~\cite{zhou2015upper} set some limitations to minimize the number of simulations done by Monte-Carlo in algorithm related to greedy. They demonstrate that their limitation is well-established when the sum of links over a node is less than one. They also proposed an approach called Upper Bound oriented Lazy Forward algorithm (UBLF) based on their limitation to identify $k$ nodes which have the significant impact in social network. Through experiments, they demonstrate that UBLF minimizes greater than 95\% Monte-Carlo simulations of CELF. Their approach outperforms in terms of performance 2-10 times quicker than CELF when the seed size is small.

Nguyen et al.~\cite{nguyen2013budgeted} developed a prototype named influence maximization of the budgeted (BIM) to choose a number of seeds impacted on social networks in value of cost in the domain of budget. Their developed model should compute the influence propagation of nominee seed which is \#P-hard. Through experiments utilizing large social networks and synthetic data, they show that their developed algorithm obtain great performance with average cost of calculation.

Tong et al.~\cite{tong2017adaptive} investigated the ways which seed users can be selected adaptively. They first made a prototype for the dynamic independent cascade prototype and then scrutinized into strategy of selection seeds adaptively. Afterwards, according to the developed model, they found out that their algorithm can choose the best answer which is efficient and can prove guarantee of efficiency. Besides this algorithm, they also provided another heuristic algorithm which is so scalable. Their outputs revealed the supremacy of the strategies of selecting seeds compared to other approaches.

Rahimkhani et al.~\cite{rahimkhani2015fast} presented a novel algorithm in terms of the linear threshold model of impact maximization. Their algorithm minimizes numbers of exploring nodes without reducing the standard to diminish its running time. Their algorithm run on two datasets and its results demonstrated that their algorithm is much quicker and more efficient in comparison with the most recent algorithms.
Wang et al.~\cite{wang2010community} developed a new algorithm named Group-oriented Greedy algorithm for identifying $K$ dominant nodes.  Their algorithm has two components:
i)~a procedure to choose groups in a social network by considering information spread; and
ii) an algorithm for choosing groups to identify influential nodes. Their experiments on large mobile datasets illustrated that their algorithm is much faster compared with the most recent greedy algorithm to identify K dominant nodes.

\textbf{Discussion}
As the literature shows, for finding seeds (or influential users in a social network), researchers have used two main stochastic diffusion models (Independent Cascade Model and Linear Threshold Model) which will be depicted in detail in Section~\ref{Main Progressive Models}. Independent Cascade model is useful for describing simple contagions where activations may be changed from a single source, such as choosing of information or viruses. However, there are several situations in which exposure to multiple independent sources are needed for a person to change his behavior. For instance, when choosing a new technology, or choosing an expensive product, people may need positive augmentation from several independent sources among their friends before taking an action, here we use linear threshold model. Besides using stochastic diffusion models, social scientists have used heuristic algorithms like greedy algorithm alongside Simulated Annealing (SA) technique to reduce the running time in influence maximization problem. Some researchers have tried to optimize greedy algorithm by exerting some changes. Finally, 
a solution would be to merge heuristic algorithms by optimizing them alongside clustering approaches in order to narrow down the number of candidates to be selected as seeds in a social network and enhancing the execution time.

\subsection{Influence Maximization: Scalability }
\label{Scalability}

There are many schemes which have been developed to address the scalability of impact maximization problem. Many related works used MapReduce~\cite{dean2008mapreduce} for processing 
large datasets and proposed distributed and parallel algorithms to analyze the social influence on large networks. Karloff et al.~\cite{karloff2010model} proposed a computation model using the MapReduce paradigm. Their model restricts numbers of machines and memory for every machine so as to have a linear output. Besides, their model permits machines to run in a polynomial time to perform computations sequentially.

Chierichetti et al.~\cite{chierichetti2010max} proposed an algorithm in terms of MapReduce programming which has the same approximation as greedy algorithm. Their results on five big datasets demonstrate that their algorithm is empirical and obtains better running time in comparison with greedy algorithm.
Lattanzi et al.~\cite{lattanzi2011filtering} developed an algorithm in terms of MapReduce called filtering. The main contribution of them was minimizing the input size in a scattered system so that each instance of the problem can be resolved on a stand-alone machine. They used different methods to show balance among the memory and the number of MapReduce rounds. Although the machines are given sub-linear memory, the rounds of MapReduce for running their algorithm is a constant. Moreover, their algorithm enjoys a significant speedup.

Kumar et al.~\cite{kumar2015fast} showed that a large class of greedy algorithms can be solved in MapReduce in the domain of submodular maximization.
They presented a technique of sampling called Sample and Prune which is so robust for Model of MapReduce. This technique tries to identify a nominee answer to the problem among an example of the input and then uses this existing answer to trim the features which cannot lead to the answer. By iterating this step, they could reach the optimal result.

Mirzasoleiman et al.~\cite{mirzasoleiman2013distributed} presented an algorithm named GREEDI in the domain of distributed submodular maximization. It was easily implemented in MapReduce computation models~\cite{dean2008mapreduce}. They showed that for problems like active selection of set and exemplar-oriented clustering, their method gave rise to parallel solutions which outperformed than those which were obtained by centralized methods.
Lucier et al.~\cite{lucier2015influence} described a new sampling approach which is for designing scalable algorithms. These algorithms can be implemented by computing frameworks such as MapReduce. They used a probabilistic analysis to explain how an algorithm can set variables to have a balance between simulation cost and informativeness. Their experiments illustrate the performance of their prototype.

Goyal et al.~\cite{goyal2011data} presented a novel model called credit distribution which used diffusion traces to be taught how impact exists in the network and also employs this to obtain impact propagation. Their method is time-oriented. They also demonstrated that impact maximization is NP-hard by credit distribution model and the function that determines influence spread is submodular. Their results show that their model has a high accuracy in comparison with the standard approach as well as scalability and efficiency.
Jung et al.~\cite{jung2012irie} developed a novel algorithm for impact maximization called IRIE that has both merits of ranking of influence (IR) and estimation of influence(IE) functions at the same time that mingles the diffusion of negative opinion. By their experiments, they showed that IRIE is more scalable than other algorithms. Besides, they proved that their algorithm is much stable based on execution time and memory usage compared with other algorithms for different size of networks.

Wang et al.~\cite{wang2012scalable} designed a novel heuristic algorithm which is able to be expanded to a vast number of edges and nodes in a social network. Their algorithm has a variable for controlling the tradeoff among the impact propagation of the algorithm and the running time. Their outputs demonstrate that their algorithm is much scalable for million-sized graphs and also their algorithm substantially outperforms in influence propagation compared to other scalable heuristics.
Heidari et al.~\cite{heidari2015smg} worked on a novel algorithm to boost the scalability of greedy algorithm. Therefore, they developed an algorithm named State Machine Greedy which enhances the most recent algorithms through decreasing computation in two sections: 1) calculating the nodes which are in diffusion function, 2) graph of Monte-Carlo in propagation simulation. Their results revealed that their algorithm can outperform in the velocity in comparison with other greedy methods.

Although there are greedy algorithms which can give good approximation to optimum output, they also have some shortcomings regarding low efficiency and long running time. Thus, Liu et al.~\cite{liu2014imgpu} presented a new structure named IMGPU to increase the proliferation of influence by using parallel processing ability of graphics processing unit (GPU). They initially enhanced the current greedy algorithms and designed an algorithm with GPU consistency, which already had parallelism. In order to set their algorithm to the architecture of the GPU, they presented a method based on k-level merging to increase the parallelism and change the graph of influence to diminish the possible variance. Finally, they run their experiments on different datasets including real-world and artificial networks. Their experimentations demonstrated that their algorithm can outperform compared to other algorithms in maximization of influence up to 60\% and also it is scalable for large networks.

\textbf{Discussion}
According to the literature and considering that large amount of information generating every second on online social networks, scalable algorithms are vital to analyze the influence maximization. Many related works used MapReduce~\cite{dean2008mapreduce} for processing and generating large datasets and proposed distributed and parallel algorithms to analyze the social influence on large networks. Besides, some researchers have used scalable sampling (using MapReduce) to achieve the optimal result in a real time manner.
Furthermore, some social scientists have tried to use heuristic algorithms (e.g., greedy algorithm) and enhance the scalability of these algorithms to obtain the better performance.
Overall, MapReduce is an asset for processing parallelizable problems through big datasets using a vast number of computers. MapReduce by using heuristic algorithms (e.g., improved greedy algorithm) can lead to scalable approach for discovering seeds in a social network efficiently.


\section{Stochastic Diffusion Models}
\label{Stochastic Diffusion Models}

Stochastic diffusion models play an important role in understanding and modeling the propagation of information. Therefore,  here we have surveyed the most studied models of propagation (Independent Cascade and Linear Threshold).
Related works model a social network as a graph $G = (V, E)$ in which $V$ is a number of nodes, and $E \subseteq (V \times V)$ is a number of ordered pairs called edges. A node indicates a person in the social network, while an edge from u to v represents the correlation between individuals v and u. They also suppose the correlation between nodes are directed and non-symmetric and they consider $G = (V, E)$ as the social graph. There are some terminologies relating to graph theory, such as a node u's outgoing $arc(u,v) \in E$ or incoming $arc(u,v) \in E$, or u's out-neighbors $N ^ {out}(u)$ or in-neighbors $N ^ {in}(u)$.

The spreading of information proceeds in distinct time steps (with time $t$= 0,1,2,...).
Each node $v$ which is in $V$ has two states, inactive and active. When a node $u$ accepts the new idea and information being spread through the network, we consider it as an active node. However, when a node has not accepted the new idea or information, it means that it is an inactive node. Let $S_t \subseteq V$ be the number of nodes which are active at time t, referred to as the active set at time t. We consider $S_0$ the \emph{seed set} and nodes in this set the \emph{seeds} of influence diffusion. Such seed nodes are the beginning nodes selected to spread the influence, for instance, the beginning users chosen by a firm to obtain free samples of the firm's new product.

\textbf{Stochastic Diffusion Model.}

A stochastic diffusion model~\cite{chen2013information}(with instinct time steps) for a social graph $G = (V, E)$ identifies the randomized procedure of producing active sets $S_t$ for all $t >= 1$ with the beginning seed set $S_0$. Here $S_t$ is a random set with the distribution stated by the stochastic diffusion model. For understanding the model of stochastic diffusion, we present an equivalent diffusion model that gives more detail about the model.

\textbf{Model equivalence.}

We consider two stochastic diffusion modes identical if for any seed set $S_0 \subseteq V$ , for any time $ t>=1 $ and any subsets $B_1, ... , B_{t-1}$, the probability for the event $\{S_1= B_1, ..., S_{t-1} = B_{t-1}\}$ is either zero or non-zero in both models.

For a class of spreading models, when a node gets active, it remains active for all $t >= 1$, $S_{t-1} \subseteq S_t$, and we name these models progressive models. However, models in which nodes may change frequently between inactive and active states are named non-progressive models. Progressive models are applied to model the spread of the choosing of state-of-the-art technologies, new products, etc. For example, choosing a new mobile phone. Nonetheless, Non-progressive models generally model the spread of ideas and opinions, opinions about a news or being fan of a political view for instance, which may change in terms of new information collected from the network.
\subsection{Main Progressive Models}
\label{Main Progressive Models}

In progressive diffusion models, because active sets do not decrease and the full set V is limited, within a limited number of steps the active set does not alter. We name active set \emph{final active set} and call it as $\Phi(S_0)$ , where $S_0$ is the beginning seed set. We suppose $\Phi(S_0)$ is a random set determined by the stochastic process of diffusion model. One main challenge studied in our literature review is to increase the anticipated number of the ultimate active set. Let $A(X)$ indicate the anticipated value of a random variable X. For this aim, we $\sigma(S_0) = A(|\Phi(S_0)|)$ and name it the influence diffusion of seed set $S_0$~\cite{chen2013information}.

In this section, we give more details about two progressive models, the independent cascade model and the linear threshold model, both of which were studied in the context of mathematics. We scrutinize their features such as the main submodularity feature.

\subsubsection{Independent Cascade Model.}
\label{Independent Cascade Model}

Independent Cascade model (or IC model) is first studied by Kempe et al.~\cite{kempe2003maximizing}, in terms of models in communication particle system~\cite{durrett1988lecture,liggett2012interacting} and marketing research~\cite{goldenberg2001using,goldenberg2001talk}. The IC model is associated with epidemic model~\cite{anderson1992infectious}. One of the main properties of this model is that the spread events alongside each edge in the social graph are independent.

In the IC model, every edge (u,v) has an impact possibility p(u,v) which is in range of [0,1] correlating with which node u affects node v. In below, we will explain the IC model technically.

\emph{Independent Cascade Model.}  The IC model~\cite{chen2013information} uses the social graph $G = (V, E)$, the impact possibility p(.) on all edges, and the beginning seed set $S_0$ as the input, and produces the active sets $S_t$ for all $t >= 1$ by the upcoming randomized operation rule. At every time step t which is greater than 1, initially set $S_t$ to be $S_{t-1}$; afterwards, for each non-active node $v \notin S_{t-1}$, for each node $u \in N ^ {in}(v) \cap (S_{t-1} \setminus S_{t-2})$, u runs an activation effort by running a Bernoulli test with possibility of accomplishment p(u,v); if doing well we insert v into $S_t$ and we say u activates v at time t. If numerous nodes activate v at time t, the ultimate result is the same --- v is added to $S_t$.\\
\subsubsection{Linear Threshold Model.}
\label{Linear Threshold model}
As we mentioned before, the IC model is appropriate for describing simple contagions where activations may be changed from a single source, such as the choosing of information or viruses. Nonetheless, there are several situations in which exposure to multiple independent sources are needed for a person to alter his behavior. For example, when choosing a new technology or a commercial idea, or choosing an expensive product, people may need positive augmentation from several independent sources among their friends before taking an action.

Social scientists have presented threshold behaviors to model such diffusions~\cite{granovetter1978threshold,schelling2006micromotives}. When a total function (e.g., sum, or count) of all of the positive signals which are received by a target is more than a determined threshold, the target is activated. Centola et al.~\cite{centola2007complex} mentioned threshold behaviors in which a person takes an action just after obtaining influence from two or more sources as complex contagion.

The linear threshold model (or LT model) is a stochastic diffusion model presented by Kempe et al.~\cite{kempe2003maximizing} to show this type of behavior. In the LT model, every edge (u,v) which is in E is related to an impact weight w(u,v) which is in range of [0,1], showing the significance of u on having an influence on v. The weights are homogenized such that for all v, the total of wights of all entering edges of v is not more than 1. For convenience, we set $w(u,v) = 0$ for all (u,v) which are not in E. In below, the formal definition of the LT model is given.

\emph{Linear Threshold Model.} The linear threshold(LT) model~\cite{chen2013information} uses the social graph $G = (V, E)$, the impact weight w(.) on all edges, and the seed set $S_0$ as the input, and produces the active sets $S_t$ for all t which is greater than 1 by the following randomized operation rule. Each node v which is in V alone chooses a threshold $ \theta_v$ randomly in the span [0,1]. At every time step $ t >= 1$, initially we set $S_t$ to be $S_{t-1}$; afterwards, for any non-active node v which is in $V \setminus S_{t-1}$, if the whole weight of the edges from its active in-neighbors is at the minimum $ \theta_v$, then insert v into $S_t$.\\

\subsubsection{Submodularity and Monotonicity of Influence Spread Function.}
\label{Submodularity And Monotonicity}
Both the IC model and the LT model have two main features based on their influence diffusion function $\sigma(.)$: monotonicity and submodularity.

Modularity can be realized in this area as reducing marginal returns when inserting more nodes to the seed set. A set function g  is submodular if for any subsets $ P \subseteq Q \subseteq W $ and any component $w \in W \setminus Q$, the marginal profit of component w when inserted to a set Q cannot be more than the marginal profit of component w when added to a subset $ P \subseteq Q $.
Formally,
\begin{equation}
g(P \cup \{w\}) - g(P) >= g(Q \cup \{w\}) - g(Q).
\end{equation}

Another important feature is monotonicity, which in this area signifies that inserting more elements to a seed cannot decrease the size of the ultimate set. We can say that a set function g is monotone if for any subsets $ P \subseteq Q \subseteq W $, $ g(P) <= g(Q) $.

\section{Social Network Analytics}
\label{Social Network Analytics}

Social networks have been investigated substantially in the general backgrounds of studying communications among individuals, and finding the significant systemic paradigms in such communications~\cite{Aggarwal}. There are two main dimensions in analyzing social networks:
content (the content such as comments and posts generated by social actors on online social networks) and
actions (activities of social actors such as liking a post, viewing a profile and following another social actor).


\subsection{Content Analysis in a Social Network}
\label{Content Analysis in a social network}

Here, we investigate some approaches which help us analyze a content of a text in a social network - such as a text of a Tweet in Twitter or a post in Facebook. Beheshti et al.~\cite{DataSynapse,AminVLDB,beheshti2017automating,DBLP:conf/caise/BeheshtiVBT18,DBLP:journals/corr/BeheshtiTBN16} proposed a framework for data curation feature. They used natural language processing technology and machine learning algorithms to change unprocessed data into processed knowledge and data., e.g. by deriving various features (keywords, topics, phrases, part-of-speech, name entities etc.) from the text of the tweet as well as URLs, images, descriptions and more in the sample tweet. Then, they automatically link the extracted items (e.g. a named entity such as a location) to knowledge bases (KBs) such as Wikidata (wikidata.org/). They also generate related and similar items such as synonyms and stems for similar keywords (e.g. leveraging Word2vec\footnote{https://www.tensorflow.org/tutorials/word2vec}). We call all the mentioned procedures "Data Curation" - transforming unprocessed data into curated data, i.e. processed knowledge and data which is maintained and available to use by end-users. In the following, we describe data curation process in detail.

\subsubsection{Extracting Knowledge From Raw Data}
A growing number of information is becoming widespread by means of text documents, web pages, emails, news and articles demonstrated in natural language text. Such data are named unstructured compared to structured data which is returned to normal and saved in a database. The scope of Information Extraction (IE) is related to recognizing information in data without any structure. Mostly, this activity relates to analyzing texts of human language by way of Natural Language Processing(NLP)~\cite{BeheshtiBVRMW17,DBLP:conf/icsoc/SchiliroBGABYSC18,DBLP:conf/icsoc/AmouzgarBGBYS18,DBLP:journals/corr/BeheshtiVRBW13}. As a result, analysts may need having access to APIs related to natural language processing to get hold of entities, keywords, part of speech and more.

\textbf{Named Entity}
A named entity is an object in the real world, like places, firms, individuals, products, etc., which could be designated by a certain name. For example, Donald Trump , Australia can be named entities.
There are some techniques regarding Named Entity Recognition (NER) in order to detect and categorize each items in text into predetermined classes like the names of firms, individuals, places, times and etc.~\cite{BeheshtiBVRMW17}. Especially, named entities have significant information regarding the text, and therefore are significant for extraction. As a result, NER is an important section of information taking out systems which provides appropriate names necessary for a large number of applications, helps pre-processing for various categorization levels, and makes information linkage easy. There are some tools which can assess named entity extraction like Stanford-NLP\footnote{http://nlp.stanford.edu/.}, OpenNLP\footnote{http://opennlp.apache.org/.}, LingPipe\footnote{http://alias-i.com/lingpipe/.}, Supersense Tagger\footnote{http://sites.google.com/site/massiciara/.}, AFNER\footnote{http://afner.sourceforge.net/afner.html.}, and Alchemy API\footnote{http://www.alchemyapi.com/.}. The descriptions of these tools are mentioned in Table~2.1.
\begin{table}
\caption{Named Entity Extraction Tools}
\label{table1}
\begin{tabular}{|l|l|}
\hline
Tools &  Description\\
\hline
Stanford-NLP &  This tool is a coherent set of NLP tools in Java programming language, \\
                    & consisting of tagging part-of-speech, recognition of named entity,  \\
                    & coreference and parsing. This tool also gives a common \\
                    & implementation of conditional random field (CRF) models which \\
                    & are accompanied by other property extractors for recognition of named entity \\
                    & \\
OpenNLP &   This tool is based on machine learning approaches for analyzing the content of \\
        & a text. It contains the most important NLP tasks like tagging of part-of-speech,\\
        & separating part-of-speech, taking named entity out,\\
        & extraction of named entity, parsing, tokenization and coreference resolution\\
        & \\
LingPipe & This tool is applied to identify named entities in mass media, categorize the search \\
         & outcomes of Twitter into groups, and recommend appropriate queries. It consists of \\
         & training with new data as well as various models such as multi-genre,\\
         & multi-lingual, and multi-domain models\\
         & \\
Supersense Tagger & This tool has been designed for tagging of verbs \\
                 & and nouns in terms of WordNet classifications which consist of organizations, \\
                 & persons, locations and quantities~\cite{ciaramita2006broad}. It is a kind of automatic\\
                 & learning, providing three various models for application: WSJ, CONLL  \\
                 & and WNSS\\
                 & \\
AFNER & This tool is open-source which is able to identify names of \\
        & people, locations of organizations , and financial quantities in \\
        & English texts~\cite{van2007named}. This tool utilizes rational expressions to discover plain named\\
        & entities. It can also detect text matching based on named entities\\
        & \\

AlchemyAPI & AlchemyAPI offers APIs to use machine learning approaches and NLP technology. \\
            & It contains various languages and provides complete solutions \\
            & having disambiguation abilities \\
            & \\
\hline
\end{tabular}
\end{table}


\textbf{Part of Speech}
A part of speech is a group of vocabularies which have identical grammatical properties~\cite{jurafsky2014speech}. The words which have identical part-of-speech indicate similar behavior based on syntax and they also follow similar grammatical structure. Some of the most prevalent part-of-speech are verb, noun, pronoun, adjective, adverb, interjection , preposition, conjunction and articles.

\textbf{Keyword}
A keyword is a vocabulary which repeats in a passage repeatedly than we would anticipate to happen accidentally per se~\cite{BeheshtiBVRMW17}. Keywords are determined by performing a statistical experiment which measures the vocabulary frequencies in a context in opposition to their anticipated frequencies obtained in a larger collection. In order to help analysts index and filter open data, it would be significant to take out keywords from semi-structured or unstructured data like tweets text.

\textbf{Synonym}
A synonym is a phrase which conveys the same or almost the same meaning as another phrase in the identical language. For instance, synonyms can be the words finish, end and stop. Two vocabularies or phrases can be synonym if they mean in determined contexts~\cite{manning2014stanford}. It is important to to take out synonyms for keywords and take them into account when analyzing the open data. For example, occasionally some tweets can be equivalent if we consider the synonyms of keywords in our analysis rather than only concentrating on the precise keyword.

\textbf{Stem}
A stem is a form which is attached to a suffix or prefix to make a new word with almost the same meaning. For instance, the word relationship contains the stem relation, to which the postfix -ship is attached to create a new stem relationship. In order to help analysts realize and analyze the textual context, it is worth extracting stems of the words in the context. If we consider the keyword `mean', applying the stem service, it would be feasible to find derived words such as meaning, meaningful, meaningfully, meaningless, etc.

\subsubsection{Linking Extracted Items to External Knowledge Bases}

Linking or Data matching generally depends on the usage of similarity function, where a similarity function $f(x_1,x_2) \rightarrow s$ is used to allocate a grade s to a couple of values $x_1$ and $x_2$. If s exceeds a given threshold t, these values are taken into account as the same real world objects. There are many similarity functions which have been presented for comparing~\cite{BeheshtiBVRMW17}: numeric values, strings , images , date and etc. Examples for each similarity function are describes as follows:

For string data types, one of the examples can be edit distance which is computed by the smallest number of edit operations( like insertion and deletions) ro change one string to another one. Regarding numeric data, one similarity function treats numbers like string and then compares them with each other like similarity functions for strings. Regarding date and time, date ought to convert to a common format so that dates can be compared together. For instance, one of common formats for date type is `ddmmyy' which `dd', `mm' and `yy' indicate day, month and year respectively. In this data type, times should be in the form of strings in 24-hour format.

\textbf{Knowledge Bases}
While extracting different properties like named entities, synonyms, keywords and stems from passage, it would be better to relate the extracted information to entities in the Knowledge Graphs available such as Wikidata and Google KG. For instance, considering extracted `B. Obama' from a tweet text. It would be feasible to find identical entity (e.g. `Barack Obama') in the Wikidata. As we mentioned before, the similarity API enjoys several functions. For these two strings, a similarity function named the Jaro function gives 0.75 and the Soundex function gives 1. To obtain this, we have applied the Wikidata APIs and Google KG to relate the extracted entities from the passage to the entities in these knowledge bases.

\subsection{Social activity Analysis in a Social Network}

There are a set of studies which concentrate on extracting different features for assessing the actions and reactions done by social actors; such as Followers-Count\footnote{Followers-Count is the number of followers which each user has in a social network.}, Follower-Ratio\footnote{
Follower-Ratio is a criterion which shows the relationship of a person's followers to its following in terms of quantity in a social network like Twitter. A positive follower-ratio means that your followers are more than the people you have followed in Twitter. A negative ratio is vice versa and it happens when your followings are more than your followers.}, Friends-Count\footnote{Friends-Count is the number of friends that each person has in a social network.}, Pageview-Count\footnote{Pageview-Count is the number of people who have seen your page in a specific time or day.}, Audience-Size\footnote{Audience-Size is the number of visitors who have visited your page, ads and status in a social network like Facebook.}, Replies-Count\footnote{Replies-Count is the number of reply messages that each user send in a social network.}, Likes-Count\footnote{Likes-Count, in Facebook for instance, is the number of likes which each user receives from his/her friends.}, Mentions-Count\footnote{Mentions-Count, is the number of people who are mentioned in each post or tag in Facebook or other social networks.}, Activity-Rate\footnote{Activity-Rate, as an example, in Facebook, the number of log-in (as an activity) that each user does in a period of time is counted as log-in rate.}, Klout-Score~\cite{rao2015klout}\footnote{Klout-Score~\cite{rao2015klout} is a criterion for assessing influence of users on social platforms like social networks and online forums. These days, Klout score is mostly used for recognizing influential users for applications like influencer marketing~\cite{schaefer2012return}.}, Profile-Rank~\cite{eirinaki2012identification}\footnote{Profile-Rank is a metric that utilizes activity traits and popularity of each user to classify them based on their impact.}, Influence-Network~\cite{tinati2012identifying}, Outreach-Network~\cite{bartholomay2011mapping}\footnote{Outreach is an activity of giving facilities to any individuals who may not access to those facilities. As an example for social network outreach, Facebook encourages social clubs to use outreach tools to engage with their followers.}, Influence/Outreach-Score\footnote{Regarding Influence/Outreach-Score, Influence score can be assessed by the number of reactions that a user can induce over time~\cite{rao2015klout} such that influential users can be compared by this score in the network.} and Influential-Followers~\cite{weng2010twitterrank,DBLP:conf/wise/BeheshtiBNA12}.

There are basic approaches which focus on network properties such as Followers-Count, Pageview-Count, Audience-Size, Likes-Count, Mentions-Count, etc. Besides, there are algorithmic approaches which try to analyze the evolution of combination of network attributes over time. These algorithms include Klout-Score, Profile-Rank, Influence-Network, Outreach-Network and Influence/Outreach-Score.
For example, for calculating Klout-Score, Rao et al.~\cite{rao2015klout} presented a framework for producing an influence score for every user. In this approach, they collected more than 3600 features through several dimensions for every user to receive signals of influential communications. In their proposed approach, they used supervised models which were trained by labeled data to compute weights for features and finally, klout score for each user was calculated. They showed that users by high klout score can diffuse information in a network more effectively.

\section{Social network analysis of actors' behavior}
\label{Radicalization}
Radicalization issue has become one of the burgeoning, new research topics and applications in the area of influence maximization on social networks. Radicalization~\cite{kruglanski2014psychology}, the process of getting involved in activities which are considered as violation of various norms. There are some work in this area which includes various research domains such as social science, computer science and psychology. These existing works have concentrated on understanding the route to radicalization: 1) choosing main signifiers of radicalized behavior~\cite{bartlett2012edge}(e.g. spread of jihad video), 2) explaining models of radicalization~\cite{king2011radicalization} (e.g. anger, isolation, disillusionment, etc.) and 3) the procedure by which radicalized individuals employ others~\cite{berger2015tailored,hall2015canadian,winter2015documenting,DBLP:journals/compsec/AllahbakhshIBBFB14,DBLP:conf/apweb/AllahbakhshIBBBF13,DBLP:conf/colcom/AllahbakhshIBBBF12}.

In addition to a few investigations of how radicalization occurs, predicting who will probably be radicalized has been the focal point of a few studies. For instance, O'Callaghan et al.~\cite{o2014online} grouped Twitter clients gathered from Twitter records associated with the Syria war into high-intensity cluster. Afterwards, they detected a group of users, which incorporated the individuals who were fan of ISIS. Checking the video which were shared by the clients in that group revealed that recordings were shared from Youtube channel supported by ISIS and Aleppo (a city which has been under control of ISIS). Besides, Berger et al.~\cite{berger2015isis} gathered a large number of supporters from Twitter manually and afterwards created a machine learning model to discriminate between proponents and opponents of ISIS supporters. By this technique, the authors discovered that proponents of ISIS can be precisely foreseen (around $94\%$ precision) via their profile explanations per se: using keywords like Islamic State, Caliphate State, Iraq and etc. Likewise, Magdy et al.~\cite{magdy2015failedrevolutions} could distinguish proponents and opponents of ISIS users, realizing that ISIS proponents chat more about the Arab Spring compared to ISIS antagonists. Two kinds of users were defined from Arabic Tweets. Proponents of ISIS used `Islamic State' more frequently while opponents of ISIS used `ISIS' more. On the other hand, Saif et al.~\cite{saif2017semantic}, investigated the semantics' role in categorizing users as proponents or opponents of ISIS - in contrast with using phrases in users' profile explanations~\cite{berger2015isis} or phrases per se~\cite{magdy2015failedrevolutions}. Their results validate the efficacy of semantics empirically (i,e. entities and relations, semantic notions).

\chapter{Proposed Model}
\label{chap:Proposed Model}


The rise of cyber-space as a new battleground for information warfare, along with the confluence of technological and societal advances is changing the nature of global terrorism.
Engagement with Web, social media, and smart devices has the potential to affect the mental behavior of the individuals and influence on extremist  and criminal behaviors such as cyber-bullying and radicalization.
Actions of violent extremists - such as subsequent attacks in Paris, London and Madrid - threaten humanities' principles and core values which include the rule of law, human rights and democracy.
While radical thinking itself is not problematic, it poses a threat to national security when it leads to violence.

It would be very challenging to understand the patterns of extremist and criminal activities on social networks, nonetheless, most of the people who have become radicalized share some common elements in their experiences, irrespective of their motivations and beliefs~\cite{radical}.
Radicalization~\cite{kruglanski2014psychology} - ``the process by which an individual, or group comes to adopt increasingly extreme political, social, or religious ideals and aspirations" - contains two main dimensions:
(i)~Personality: the motivational and ideological component that sets a goal to which one may be committed and finds the means of violence; and
(ii)~Influence: the social process of networking and group dynamics, where cyber-enabled propaganda campaigns are increasingly used to exert strategic influence.

Social networks analytics have been studied considerably in the general context of analyzing interactions among people, and to determine the main structural patterns in such interactions~\cite{Aggarwal}.
However, major barriers to the effective understanding of extremist and criminal behaviors on social networks is the ability to understand the context as well as eventual influence that an influence campaign may have, given an understanding of the mode of information dissemination and indicators that identify the use of artificial promotion of a theme.
This task is challenging and requires technique for:
(i)~the discovery, interpretation, and communication of meaningful patterns in social data (i.e. social data analytics) to explore the potential for understanding and analyzing the personality dimension; and
(ii)~finding small subset of nodes in a social network which can maximize the spread of influence, i.e., Influence maximization~\cite{chen2009efficient}, for understanding and analyzing the influence dimension.

In this chapter, we provide a social data analytics pipeline, namely iRadical, to allow analysts to get involved in social data to explore the potential for online radicalization. In iRadical, we consider both personality and influence dimensions as first class citizens.
We introduce novel algorithms to analyze the user activity patterns to learn how influence flows in social networks.
We present the novel concept of Social Personality Graph, to model and analyze factors that are driving extremist and criminal behavior.
We implement iRadical as an extensible architecture that is publicly available on GitHub\footnote{https://github.com/DataAnalyticsResearchGroup/iRadical}. We evaluate the performance and the quality of the results using real-world and synthetic data.

Figure~3.1 illustrates the iRadical framework. 
In Section~\ref{PersonalityGraph}, we present the notion of Personality Graph and discusses how it enables analysts to understand and predict the eventual influence of a potential influencer as well as an influence campaign.

\begin{landscape}
\begin{figure} [t]
\hspace*{-1.7cm}
\centering
\includegraphics[width=1.1\linewidth]{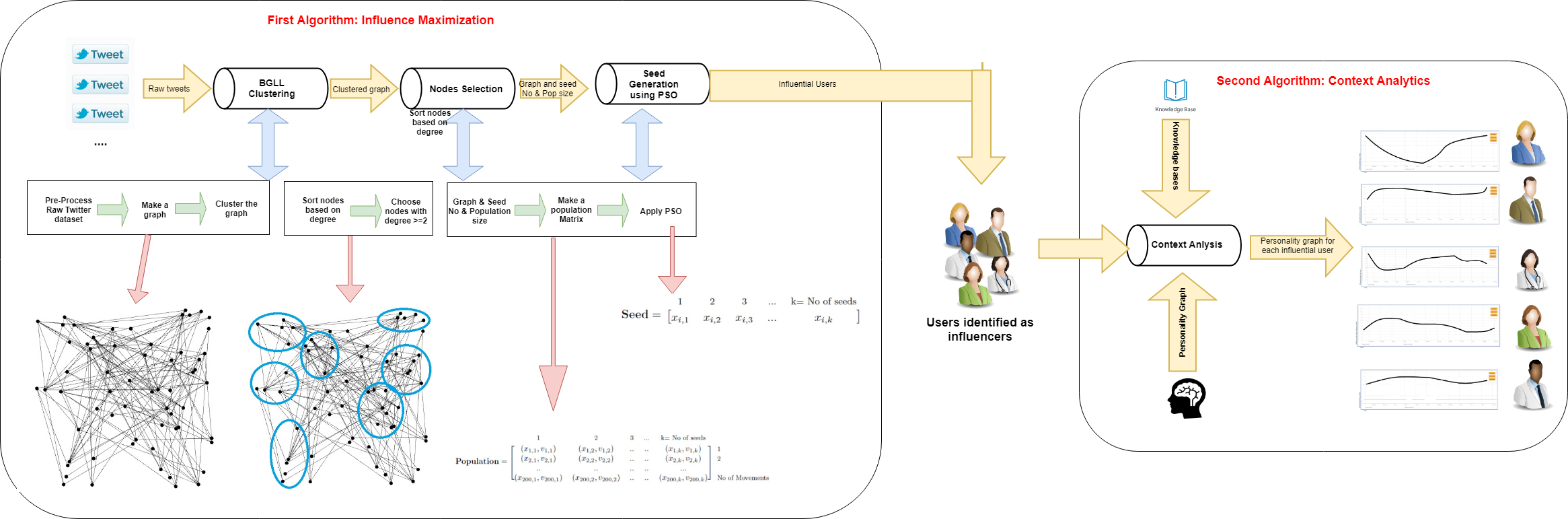}
\caption{Framework of our algorithm.}\label{fig:Figure1}
\end{figure}
\end{landscape}

\section{Constructing Social Personality Graph}
\label{PersonalityGraph}

The notion of Knowledge Lake~\cite{AminVLDB,DataSynapse,DBLP:journals/spe/BeheshtiBM18,DBLP:conf/bpm/BeheshtiSGABYSC18,DBLP:conf/cikm/BeheshtiBNCXZ17} has been introduced to provide the basis for big data analytics by curating the raw social data automatically and to process them for deriving insights.
%
For example, given a sample tweet in Twitter, the Knowledge Lake will enable analysts to automatically:
(i)~extract various features (keywords, topics, phrases, part-of-speeches, named entities) from the text of the tweet as well as URLs, images, descriptions and more in the sample tweet;
(ii)~link the extracted items (e.g. a named entity such as a person) to knowledge bases (KBs) such as Wikidata\footnote{https://www.wikidata.org/}; and
(iii)~compute related and similar items such as synonyms and stems for keywords and entities, e.g., dictionaries and lexical database such as WordNet\footnote{https://wordnet.princeton.edu/} and techniques such as  Word2vec\footnote{https://www.tensorflow.org/tutorials/word2vec}.

In this dissertation, we aim to extend the Knowledge Lake~\cite{AminVLDB,DBLP:conf/www/BeheshtiTBN17,DBLP:books/sp/BeheshtiBSGMBGR16,DBLP:conf/caise/BeheshtiBN13} to enrich social items (e.g., a tweet in Twitter) with features related to the activity of social actors. For instance, to enrich a tweet with features such as:
%
Followers-Count, Follower-Ratio, Friends-Count, Pageview-Count, Audience-Size, Replies-Count, Likes-Count, Mentions-Count, Activity-Rate, Klout-Score, Profile-Rank, Influence-Network, Outreach-Network, Influence/Outreach-Score and Influential-Followers.
Besides, we use LIWC\footnote{https://liwc.wpengine.com/} (leverage Linguistic Inquiry and Word Count) to extract psychological features from the social data, e.g. a post in Facebook, such as \emph{affective features} (e.g., positive feeling, negative feeling such as depression, anxiety and grief), \emph{Cognitive features} (e.g., insight, causation and discrepancy) and \emph{Personal obsessions features} (e.g., achievement, leisure and work).
We leverage existing algorithms and techniques to automatically extract these features and take them into consideration to construct a social personality graph for each user/actor in an online social network~\cite{DBLP:journals/cluster/BatarfiSFNBBS15,DBLP:journals/pvldb/HammoudRNBS15}. We designed a mockup model related to radicalization analysis using gold standards such as LIWC\footnote{http://liwc.wpengine.com/} to extract various features from the social data as follows in Fig~3.2.
\begin{figure} [t]
\centering
\includegraphics[width=0.9\linewidth]{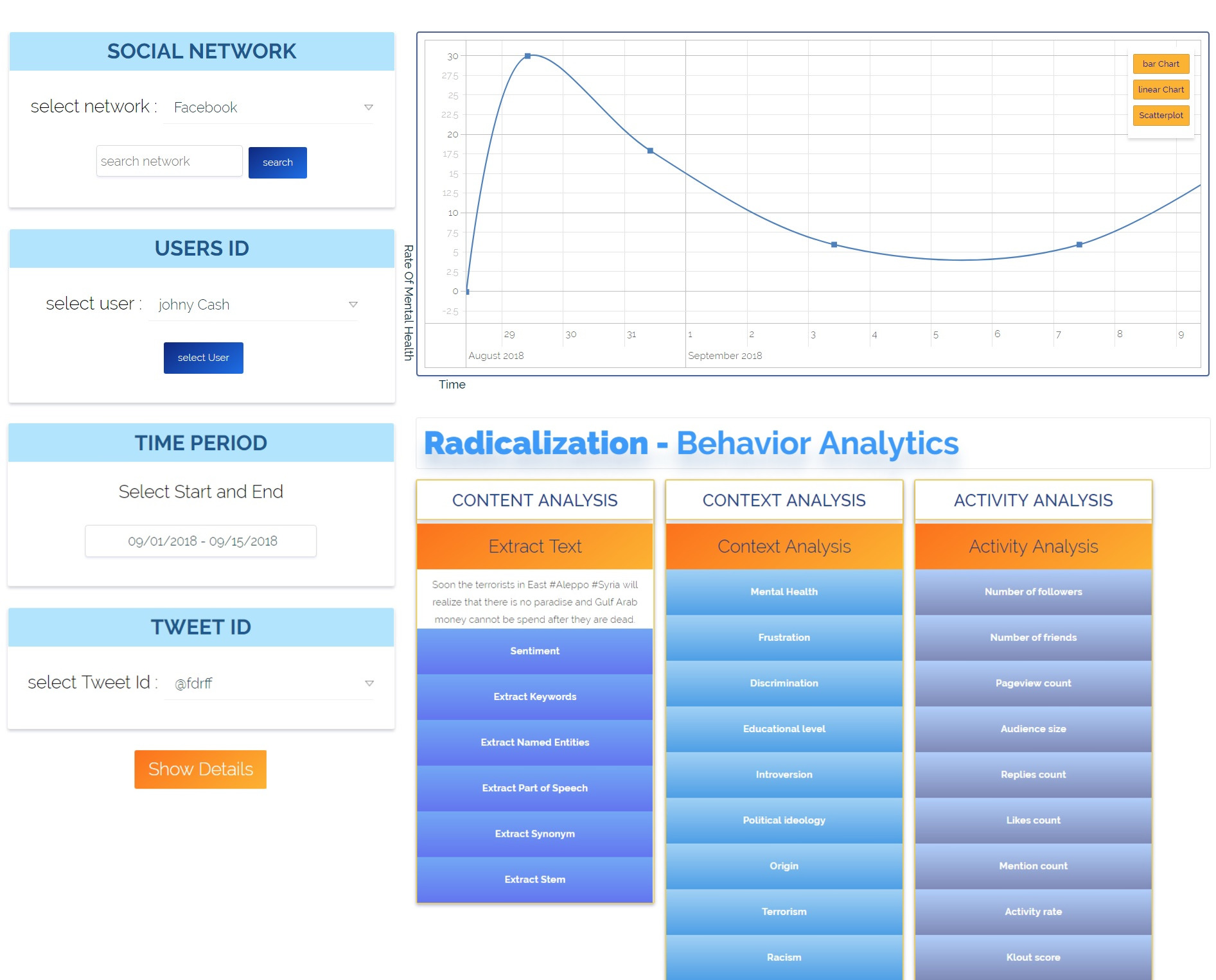}
\caption{A software prototype, implemented to analyze the users' behavioral disorder and to identify radicalized people on online social networks.}\label{fig:Demo}
\end{figure}


\emph{Definition 1. (Social Personality Graph) A social network is modeled as a graph $G = (V, E)$ in which $V$ is a number of nodes, and $E \subseteq (V \times V)$ is a number of ordered pairs called edges. A node indicates a person in the social network, while an edge from u to v represents the correlation between individuals v and u. We suppose the correlation between nodes are directed and non-symmetric  and we consider $G = (V, E)$ as the social graph. There are some terminologies relating to graph theory, such as a node u's outgoing $arc(u,v) \in E$ or incoming $arc(u,v) \in E$, or u's out-neighbors $N ^ {out}(u)$ or in-neighbors $N ^ {in}(u)$.}

\emph{Definition 2. (Entity) An entity $E$ is indicated as a data object which has a unique identity.
Entities are illustrated by a number of \emph{attributes}.
Entities can be simple such as social user (e.g., a user in Facebook) and a social item (e.g., a comment sent by a social user).
Entities can be atomic \emph{information items} such as a keyword, topic and named entity (e.g., people, location, organization) extracted from unstructured artifacts such as a Tweet in Twitter.
Entities can be complex and representing:
\begin{itemize}
  \item social data features (e.g., keywords, topics, part-of-speeches, named entities).
  \item social activity features (e.g., Followers-Count, Friends-Count and Likes-Count).
  \item affective features (e.g., negative feeling such as depression, anxiety and grief).
  \item cognitive features (e.g., insight, causation and discrepancy).
  \item personal concerns features (e.g., work, achievement and leisure).
\end{itemize}
}

\emph{Definition 3. (Relationship) A relationship can be a directed link among two entities, which is related to a predicate determined on the attributes of entities which distinguishes the relationship. Relationships are explained by a number of \emph{attributes}.
We define the following explicit relationships~\cite{DBLP:journals/dpd/BeheshtiBM16,DBLP:conf/edbt/BeheshtiBM16}:}
\begin{itemize}
  \item \textcolor{blue}{$User$ $\xrightarrow{(\emph{post [timestamp]})}$ $Item$}: express that a user (social actor) posted an item (e.g. a tweet) on a social network (e.g. Twitter) at time $\tau_i$.
  \item \textcolor{blue}{$User$ $\xrightarrow{(\emph{follow})}$ $User$}: express that a user (social actor) follows another user on a social network.
  \item \textcolor{blue}{$Item$ $\xrightarrow{(\emph{contain})}$ $social data features$}: express that an item (e.g. a tweet) contains a set of atomic entities such as keywords, topics and named entities.
  \item \textcolor{blue}{$User$ $\xrightarrow{(\emph{contain})}$ $social activity features$}: express that a user (social actor) contains a set of social activity features (e.g., Followers- and Friends-Count).
  \item \textcolor{blue}{$Item$ $\xrightarrow{(\emph{contain})}$ $affective features$}: express that an item (e.g. a tweet) contains a set of affective features (e.g., negative emotion like anxiety and anger).
  \item \textcolor{blue}{$Item$ $\xrightarrow{(\emph{contain})}$ $Cognitive features$}: express that an item (e.g. a tweet) contains a set of Cognitive features (e.g., insight, causation and discrepancy).
  \item \textcolor{blue}{$Item$ $\xrightarrow{(\emph{contain})}$ $Personal concerns features$}: express that an item (e.g. a tweet) contains a set of Personal concerns features (e.g., achievement/leisure).
\end{itemize}

The social personality graph will facilitate the discovery, interpretation, and communication of meaningful patterns in social data to explore the potential for understanding and analyzing the personality dimension.
The Personality Graph is time-aware (\textcolor{blue}{$User$ $\xrightarrow{(\emph{post [timestamp]})}$ $Item$}) and enables analysts to understand and predict the eventual influence of a potential influencer.

\section{Personality Aware Influence Maximization}
\label{The Proposed Algorithm for Influence Maximization}

In this section, we present 
two novel algorithms for Influence Maximization and Context Analytics. 
The stages of the first algorithm (Influence Maximization) include:
(i)~Dataset clustering by BGLL~\cite{blondel2008fast} algorithm;
(ii)~Nodes selection; and
(iii)~Seed generation by PSO algorithm and local search procedure.
The second algorithm (context analytics) evaluates the radicalization risk in influential users obtained from the previous algorithm.
Fig.~3.1 illustrates the framework of our algorithm.

\subsection{First Algorithm: Influence Maximization}
\label{First algorithm}

\textbf{Pre-processing.}
In this stage, we make an undirected weighted graph from the Twitter dataset provided by Kaggle~\cite{Kaggle}. The dataset includes the following: Name, Username, Description, Location, Number of followers (at the time the tweet was downloaded), Number of statuses by the user (when the tweet was downloaded), Date and timestamp of the tweet and the tweet. Each node in our graph refers to each username in the dataset. Each edge (which has a weight) between two nodes indicates pairwise co-occurrence of all words in any tweet for each username, divided by the sum of all string words from both usernames being compared. Therefore, the weight between each two nodes is a number in range of 0 to 1 which assesses the similarity of words among each two users. For example, considering that user A has 10 tweets consisting of 100 string words and user B has 20 tweets including 200 words. Moreover, the number of pairwise co-occurrence of words in all tweets for both users A and B are 45, thus, the weight between A and B will be $(45/ (100+200) = 0.15)$.

\textbf{Dataset Clustering.}
Using the proposed social personality graph data model,
we construct the personality graph for each user. This will enable us to analyze the personality of the influencers in the large graph of connected users' personality graphs.
%
To achieve this goal, we cluster all the nodes in various communities based on BGLL algorithm~\cite{blondel2008fast}.
BGLL algorithm is a clustering algorithm which was presented by Blondel et al~\cite{blondel2008fast} regarding optimization of modularity, which comprises of two stages.
At the primary stage, every node of the network is taken into account as a community. At the same time, they expel a node from its unique network to its neighbor's community that has the greatest positive gain in modularity. This stage is used over and over for all nodes till no more change can be accomplished. Afterwards, the first stage is  finished~\cite{blondel2008fast}. The communities acquired in the first stage as nodes are considered in the second stage with the goal that another network can be constructed. Next, BGLL performs these two stages repeatedly till accomplishing an unaltered outcome and acquiring the greatest modularity.

Despite other clustering algorithms, the BGLL heuristic can find more common patterns of networks since it does not need 
information regarding the number of community. Therefore, communities acquired by the BGLL get nearer to the intrinsic communities in network. In the meantime, BGLL just needs a couple of repetitions to acquire a greatest modularity, which causes the BGLL to perform better based on execution time when it is utilized to a vast-scale network.

The quality of clusters (communities) detected from this algorithm is assessed by a metric called modularity. The modularity of a community is a number in the range of -1 to 1 which assesses the frequency of links inside communities in comparison with links among communities~\cite{newman2006modularity}. Modularity can be described as Q in the networks which have weight on their links as follows.~\cite{newman2006modularity}:

\begin{equation}
Q =
\frac{1}{2\times m}\sum_{i,j}[{A_{ij}-\frac{k_i\times k_j}{2\times m}}]\delta(c_i,c_j)
\\
\end{equation}
Where $A_{ij}$ refers to the edge's weight among $i$ and $j$, $k_i=\sum_{j}{A_{ij}}$ is the weighted sum of all the edges associated with node $i$, node $i$ is attributed to the community $c_i$, $m=\frac{1}{2}\sum_{ij}{A_{ij}}$ and
\begin{equation}
\delta(u,v)= \left\{ \begin{array}{rcl}
1&
\mbox{for} & u=v \\0 & \mbox{for} &otherwise \\
\end{array}\right.
\end{equation}

\textbf{Prospective Selection.}
The goal behind this part is to recognize a set of prospective nodes as per data about communities extracted in the initial part.
Since social networks are typically greatly extensive, the search space for chosen seeds is additionally immense. Thus, we have to diminish the quantity of prospective nodes.

By examining the communities' patterns, we discover that not all the networks are sufficiently noteworthy to lodge seed nodes. For instance, in Fig. 3.3, despite the fact that the system is partitioned into three communities, community 3 might be negligible contrasted with community 1 and 2 because these have smaller size. In the event that we pick a seed node from community 3, it might just initiate three nodes at first. Therefore, we pick the community 1 and 2 as the huge communities since their size is bigger and the chance for having influential nodes in them, is high. We characterize critical communities as the principal \emph{n} big communities, where \emph{n} differs when networks change. Nevertheless, there are numerous nodes in community 1 which have common neighbors. For instance, there are 4 common neighbors of nodes $E$ and $C$, which is not helpful to influence diffusion. To tackle the bottleneck of influence which overlaps, we utilize a resemblance-based high degree function named RHD that is depicted later.

Let \emph{Prospective} be a prospective node pool. Then, a several potential nodes could be chosen out of each substantial community which are subsequently added to the \emph{prospective nodes}. The node having the highest number of neighbors is the higher degree node, which implies that node with higher degree can impact a large number of nodes with a similar spread probability. Accordingly, we choose potential nodes from every substantial community in terms of the centrality of degree. By the following formula~(3.3), we determine the number of prospective nodes opted via each substantial community~\cite{rahimkhani2015fast}:

\begin{equation}
(\frac{K_i-Min_K}{Max_K-Min_K})\times \theta + \lambda
\end{equation}
Where $K_i$ refers to the size of community $i$, the smallest community size is referred as $Min_K$ , whereas the largest is denoted as $MAX_K$. $\theta$ is the magnification term and $\lambda$ is constant which ensures the minimum number of nodes in every community. After choosing the \emph{Pospective}, the final stage of our first algorithm is to produce the final seeds.
\begin{figure} [t]
\hspace*{1cm}
\centering
\includegraphics[width=.7\linewidth]{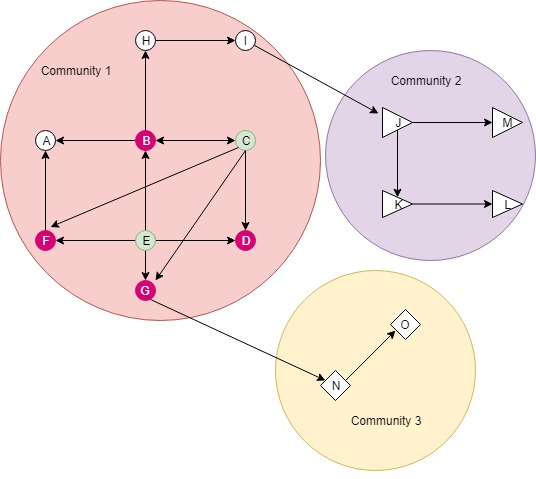}
\caption{There is a network consisting of three communities. We put label on communities as 1, 2, and 3. There are 4 red nodes in community 1 which are common neighbors of the node E and C.}\label{fig:Figure4}
\end{figure}

\textbf{Seed Generation using PSO algorithm.}
After the two described steps, the search space has been decreased. In this stage, we will use a PSO method~\cite{marini2015particle} to produce the ultimate nodes by enhancing the 2-hop influence propagation~\cite{lee2014fast} which we call our algorithm as PSO-IM. The whole framework in this step is shown in the following as Algorithm 1.

In step 1) PSO-IM fulfills the task of population initialization. To begin with, it generates the first population of solutions Pop = $\{x_1,x_2, x_3 , ...\}^T$. Afterwards, it opts the individual with the highest fitness as Popbest. Step 3 is a PSO algorithm which works as follows.\\
1) Create a matrix of population initialization. For making this matrix:\\
1.1) Initialize the position(x) and velocity (v) of each particle (node) at first.\\
1.2) Calculate fitness for each particle at this step.\\
1.3) In each movement, the history (velocity and position) of every particle is saved and the best fitness among all of his history is called $pbest_i$ (we call it personal best).\\
2) Set the number of generation to zero.\\
3) For all nodes do the following steps:\\
3.1) Update the velocity of each node by the following eq.\\
$vel_i(num_{gen}+1)=vel_i(num_{gen})+ cog_{eff}(pbest_i- x_i(num_{gen}))D_1+ soc_{coe}(gbest- x_i(num_{gen}))D_2 $ \\
Where $pbest_i$ equals personal best of the particle and gbest equals global best.\\
$cog_{eff}$ refers to ``cognitive coefficient" and $soc_{coe}$ refers to ``social coefficient" so that $cog_{eff} >=0$ , $soc_{coe} <=4$ \\
$D_1,D_2$ are diagonal matrices of random numbers having a uniform distribution such as $0=<D_1<=1$ and $0=<D_2<=1$ \\
3.2) Update the position of each node by the following eq.\\
$x_i(num_{gen}+1)=x_i(num_{gen})+ vel_i(num_{gen}+1)$\\
3.3) Particle's fitness to be evaluated as $f(x_i(num_{gen} +1))$\\
3.4) If $f(x_i(num_{gen} +1)) >= f(pbest_i)$ ,  $pbest_i$ gets updated as $x_i(num_{gen}+1)$\\
3.5) If $f(x_i(num_{gen}+1)) >= f(gbest)$ ,  gbest gets updated as $x_i(num_{gen} +1)$\\
Then in step 3.6) the localsearch algorithm is applied to find the optimal nodes among the best nodes so far. And in step 4) once the algorithm ends in convergency, PSO-IM halts and the final k-node seed is generated.

\begin{algorithm}
 \KwData {
 1) No of seed
 2) Population size (= the No of movements for particles in PSO algorithm)
 3) Maximum generation:maxgen
 4) Candidate (=No of selected nodes from the previous section) }
 \KwResult{The most influential k-node set.}
    1.Initialization;\\
        1.1 Initiate position $x_i(0) \forall i\in 1:N $;\\
        1.2 Set the particle's best position to its first position $pbest_i(0)=x_i(0)$;\\
        1.3 Estimate each particle's fitness as per formula(3.4) and if $f(x_j(0)) >= f(x_i(0)) \forall i\neq j$ the global best is initiated as $gbest=x_j(0)$ \\
    2.Initialize $num_{gen}=0$; // the number of generation \\
    3.\For{Each of N articles} {
        3.1 Update the particle velocity as per the following eq. \\
         $vel_i(num_{gen}+1)=vel_i(num_{gen})+ cog_{eff}(pbest_i- x_i(num_{gen}))D_1+ soc_{coe}(gbest- x_i(num_{gen}))D_2 $ \\
       Where $cog_{eff}$ refers to ``cognitive coefficient" and $soc_{coe}$ refers to ``social coefficient" and $cog_{eff}>=0$ , $soc_{coe}<=4$.\\
       and $D_1,D_2$ are diagonal matrices of random numbers having a uniform distribution such as $0 \leq D_1\leq 1$ and $0 \leq D_2\leq 1$\\
        3.2 Update the particle position as per below stated eq. \\
        $x_i(num_{gen}+1)=x_i(num_{gen})+ vel_i(num_{gen}+1)$\\
        3.3 Particle's fitness to be evaluated as $ f(x_i(num_{gen}+1))$.\\
        3.4 If $f(x_i(num_{gen}+1)) >= f(pbest_i)$ is true , personal best gets updated as: $pbest_i= x_i(num_{gen}+1)$ \\
        3.5 If $f(x_i(num_{gen}+1)) >= f(gbest)$ is true, global best gets updated as: $gbest=x_i(num_{gen}+1)$ \\
        3.6 Perform Local search()\;}
    4.Stopping criterion: if $num_{gen} < maxgen$, then $num_{gen}=num_{gen}+1$ and proceed to step 3) else the algorithm gets quitted and the best solution is our seed which is represented by $gbest$.
 \caption{Framework of PSO-IM}
\end{algorithm}

An elaborate illustration of numerous significant methods, i.e., fitness function(formula(3.4)), population initialization, PSO algorithm and local search algorithm are illustrated as follows~\cite{gong2016efficient}:

\textbf{Fitness function}
\begin{equation}
\Delta(X)=\sum_{x \in X}\Delta(x)-(\sum_{x \in X}\sum_{k \in K_x \bigcap X}p(x,k)(\Delta_k^1-p(k,x)))-\rho
\end{equation}
Where $K_x$ refers to the 1-hop nodes for node x (= neighbors of x),
p refers to the IC model's propagation probability, where:

\begin{equation}
\Delta_k^1=1+\sum_{k{\in}K_u} p(u,k)
\end{equation}
The term $\Delta_k^1$ refers to the 1-hop influence spread:

\begin{equation}
\rho=\sum_{x \in X}\sum_{k \in K_x \bigcap X}\sum_{d \in K_x \bigcap X -\{x\}}p(x,k)p(k,d)
\end{equation}
We divide formula (3.4) into three terms in order to explain them in detail; $\sum_{x \in X}\Delta(x)$ is denoted as term 1. $(\sum_{x \in X}\sum_{k \in K_x \bigcap X}p(x,k)(\Delta_k^1-p(k,x)))$ is term 2 and finally we consider $\rho$ as the third term.
Term 1 refers to the 2-hop influence spread for every node within S. To establish that one seed is in the neighborhood of the other seed, the redundant situation is taken into account via the second term. Lastly, in order to demonstrate that a seed is 2-hop further from another seed, the redundant situation is taken into account via the third term.

Gong et al.~\cite{7515281} proved that the 2-hop influence spread is sufficiently valid and efficient to estimate the influence spread of a
node set. Therefore, we adopt the 2-hop influence spread in our algorithm.


\textbf{Population initialization}

in PSO-IM, every individual node $X_a (1<=a<=pop)$ within the population signifies a set of seed nodes, k, which is further encoded in the following manner.\\
$X_a=\{x_a^1, x_a^2, x_a^3, ..., x_a^k\}$\\
Here, the number of seeds are depicted as k, $x_a^i$ corresponds to a node shortlisted from a group of prospective nodes. It is further noticed that $x_a$ does not possess any repeated node in itself. Let's consider that the randomly selected k nodes from the group of prospective nodes are of lower quality and subsequently transpire in a largest convergence time. In order to address the same, we employed a resemblance-based high degree heuristic (RHD) that ensures the diversity and convergence of the individual nodes. The procedure for the initialization of population is described as Algorithm 2.

In order to guarantee influence maximization on social networks (i.e., Twitter, Facebook, etc.) a standard methodology of graph theory and network analysis, i.e., higher degree of centrality, is generally employed. Nevertheless, such sort of methodology may lead to the overlapping of influence spread among the individual nodes. In order to tackle such a bottleneck, RHD has been employed. Therefore, once a node is selected, its similar neighboring nodes are excluded. Our proposed RHD heuristic then chooses the next prospective node out of all the leftover prospective nodes. The procedure is iterative in nature and continues till the requisite k nodes are selected. Structure similarity (defined in 3.7) is utilized to calculate the higher degree of similarity. The description for structure similarity is given in Algorithm 3.
In Algorithm 3, $N(v)=\{\exists u \in V,uv \in E\}$ refers to the neighbors of node v and the similarity among nodes u and v is defined as (3.7)~\cite{gong2016efficient}.
\begin{equation}
Similarity(u,v)=\frac{|NG(u)\cap NG(v)|}{|NG(u)|+|NG(v)|}
\end{equation}
Where $NG(v)=\{v|v \cup N(u)\}$ consists of both node v along with its adjacent neighbors. In algorithm~3, $0 \leq sim_{thresh} \leq 1$  and is determined based on various datasets. When the Similarity(u,v) is larger than $sim_{thresh}$, it means that u and v are similar.
The population initialization algorithm is described in Algorithm~2.

\begin{algorithm}
 \KwData{Size of population: xyz}
 \KwResult{Population Pop}
    1.50\% of population gets generated depending on RHD, refer to algorithm 3 for more detail;\\
    2.\For{n=1 to (xyz/2)} {
        \For{m=1 to k}{
            If $rand(1)> 0.5$  \\
                $x_i$ gets substituted with a selected random nodes different from each node in $x_i$;\\
            end
        }
 }
    3.\For{i=(xyz/2+1) to (xyz)}{
        $x_i$ gets initiated from k different nodes among \emph{the Prospective nodes} depending on RHD;
    }
 \caption{Population initialization}
\end{algorithm}
\begin{algorithm}
    1.Initialize $ x_a= \varnothing$;\\
    2.Temp = \emph{Prospective nodes}; \\
    3.\For{j=1 to k} {
        3.1 Select a node $v_i \in $ Temp with the highest degree;\\
        3.2 $x_a \leftarrow x_a \cup \{v_i\} $ \\
        3.3 $SimNeighbor \leftarrow \{u \in N(v)| Similarity(u,v)\geq sim_{thresh}\}$;\\
        3.4 $ Temp \leftarrow \{v|v \in Temp, v\notin SimNeighbor, v\neq v_i\}$\\
        3.5  If $Temp=\varnothing$ do\\
            3.5.1 $x_a \leftarrow x_a \cup \{v_{i+1},v_{i+2},...,v_k\}, v_{i+1},v_{i+2},v_k $ are opted from \emph{the Prospective nodes} randomly;\\
            3.5.2 Break;
 }
    4. Return $x_a$
    \\
 \caption{RHD algorithm}
\end{algorithm}

\textbf{The PSO algorithm}\\
In PSO, each prospective solution is named a ``particle" and indicates a point in a D-dimensional space, if D is the quantity of parameters to be improved. As a result, the situation of $i^{th}$ particle might be portrayed by the vector $x_i$:
\begin{equation}
x_i = [x_{i1}x_{i2}x_{i3}...x_{iD}]
\end{equation}
and the population of N prospective solutions forms the swarm.

\begin{equation}
X = \{x_1, x_2, x_3, .., x_N\}
\end{equation}

In order to find the best answer of the issue, the particles characterize directions in the parameters space (i.e., repeatedly refresh their locations.) in light of the accompanying equation of movement:
\begin{equation}
x_i(num_{gen}+1) = x_i(num_{gen}) + vel_i(num_{gen}+1)
\end{equation}
where $num_{gen}$ and $num_{gen}+1$ are two consecutive repetitions of the algorithm and $vel_i$ is the vector gathering the speed-segments of the $i^{th}$ particle through the D-measurements. The speed vectors administer the manner in which particles move over the hunt space and are made of three terms: the first, characterized the inactivity stops the particle from radically altering direction, by monitoring the past stream direction; the second term, known as the cognitive segment, represents the inclination of particles to come back to their own beforehand discovered optimal positions; the last one, known as the social segment, recognizes the penchant of a particle to move in the direction of the optimal position of the entire swarm. In terms of such contemplations, the speed of $i^{th}$ particle is characterized as:
\begin{equation}
vel_i(num_{gen}+1)=vel_i(num_{gen})+ cog_{eff}(pbest_i- x_i(num_{gen}))D_1+ soc_{coe}(gbest- x_i(num_{gen}))D_2
\end{equation}
where $pbest_i$ indicates ``personal best" of the particle that is the directions of the optimal answer got till now by that particular individual, while gbest indicates the ``global best" that is the final optimal answer got by using the swarm. The increasing speed constants $cog_{eff}$ and $soc_{coe}$, which are genuine-valued and more often in span of $0 <= cog_{eff}$ and $soc_{coe} <= 4$, refer to ``cognitive coefficient" and ``social coefficient" and regulate the extent of the stages that the particle takes toward its personal best and global best, respectively. On the flip side, $D_1$ and $D_2$ signify two diagonal matrices of irregular numbers created from a uniform diffusion in range of 0 to 1, with the goal that both the social and cognitive parts affect the speed modification rule~\cite{DBLP:conf/icse/TabebordbarB18} in Eq. (3.11). Consequently, the directions followed by the particles are semi-arbitrary inherently, as they are obtained from methodical appeal in the direction of the personal and global optimal answers and stochastic weighting of these two increasing speed terms.

All in all, the repeated procedure portrayed by Eq. (3.10) and (3.11) is rehashed till a criterion of stopping - which might be, for example, a pre-described total number of repetitions , a greatest number of repetitions because the last update of global optimal or a predetermined target estimation of the fitness is satisfied.\\
\textbf{Local Search Function}

Local search technique is an independent support strategy that is to discover a comparatively good answer among the best answers discovered till now~\cite{blum2011hybrid}. For a particle, when we alter a node to another node in the \emph{Prospective} which is not the same as another node in the particle, a neighbor of the particle is acquired. Here, we utilize a resemblance-based technique as the local search procedure. The strategy is run on a haphazard particle in the populace, at the same time it tries to discover a superior particle from the neighbors of the particle. The fitness value is computed by (3.4) and the neighbor particle is comparatively good, if its fitness value is bigger than the first particle's one. If the alteration can generate a comparatively good particle, this alteration is acknowledged. The procedure rehashes till the point when no more alteration can be done. Here, we discover the fittest particle in $P_{child}$ that is acquired after applying PSO and local search on it. The pseudocode of local search algorithm is presented as Algorithm~4.

In Algorithm 4, in the start, we run FindOptimal function to choose the particle with the greatest fitness value among the input particles. Afterwards, we utilize the local search function.
The Fitness function is to calculate the fitness of an answer in terms of (3.4). The FindOptimalNeighbour procedure is to discover the optimal adjacent particle with the highest fitness value.

\begin{algorithm}
 \KwData{$P_{child}$}
 \KwResult{$P_{child}$}
    1. $Node_{current} \leftarrow FindOptimal(P_{child})$;\\
    2. IsLocal = FALSE;\\
    3. \textbf{Repeat}\\
    4. $Node_{next}\leftarrow FindOptimalNeighbour(Node_{current})$;\\
    5. If $Fitness(Node_{next}) > Fitness(Node_{current}$) \\
    6. $Node_{current} \leftarrow Node_{next}$\\
    7. else\\
    8. IsLocal = TRUE;\\
    9. end if\\
    10. \textbf{until} IsLocal is TRUE
 \caption{The local search algorithm}
\end{algorithm}

\begin{figure}
\centering
\includegraphics[width=0.8\linewidth]{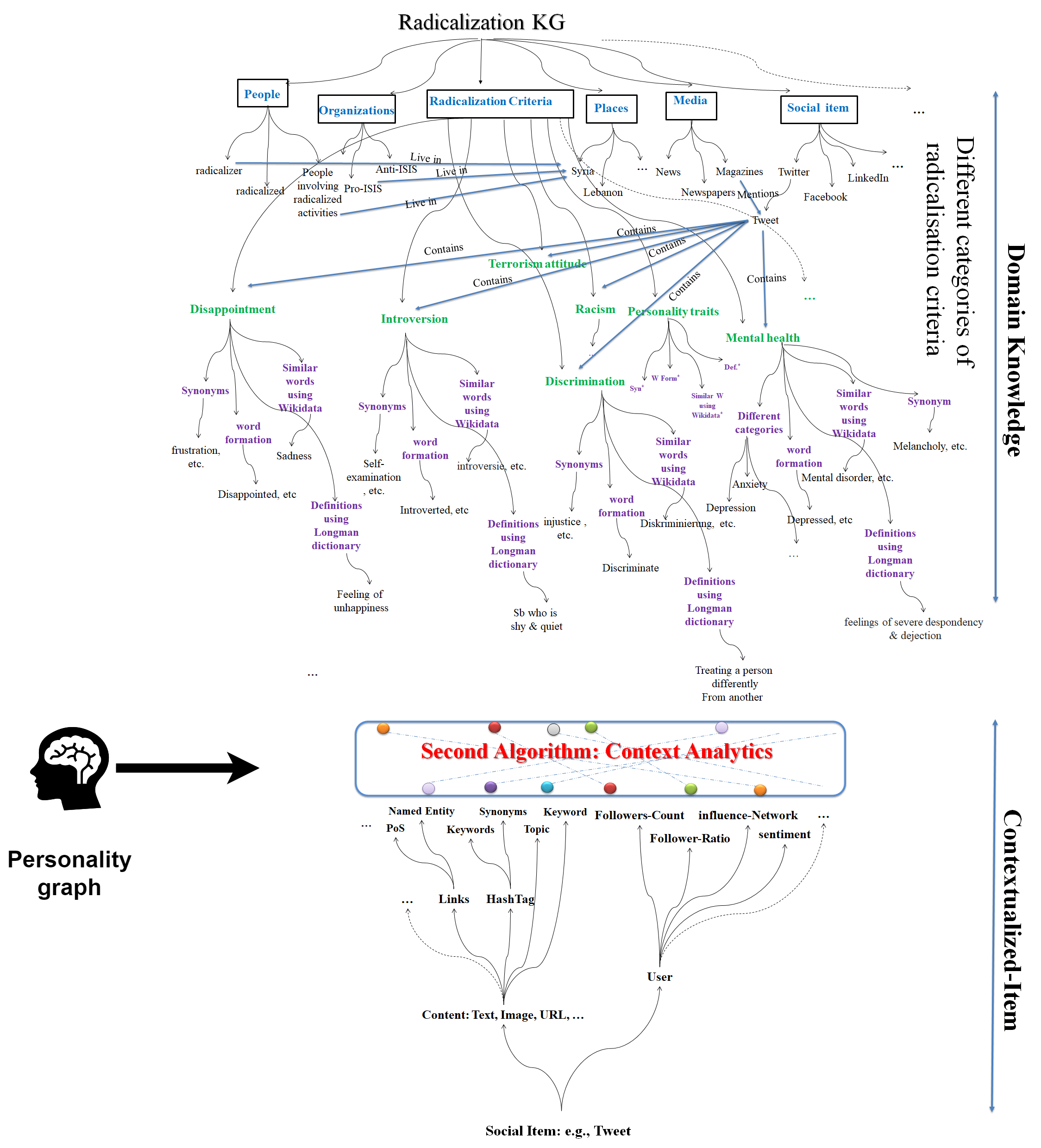}
\caption{A typical scenario for the Second Algorithm: Context Analytics.}\label{fig:Figure3}
\end{figure}

\subsection{Second Algorithm: Context Analytics}
\label{Second algorithm}

\begin{algorithm}
 \KwData {
 1) seeds from Algorithm(1)
 2) Knowledge bases
 3) Tweets of seeds }
 \KwResult{Rate of each radicalization criteria for each seed.}
    1. getwords(filepath)\\
    {
    words = []\\
    with open(filepath) as fp: \\
       line = fp.readline()\\
       while line:\\
           words.append(line.strip())\\
           line = fp.readline()\\
    return words\\
    }
    2. getrate(Text of each tweet, words of knowledge base)\\
    {
        count = 0 \\
        string wrd\\

    for wrd in words:\\
        if wrd in Text:\\
            count += 1\\
    if len(words) == 0:\\
        return 0\\
    return float(count) / float(len(words)) \\
    }
    3. for item in seeds:{ \\
        print(rate of each radicalization criteria for each seed)\\
             }
 \caption{Second Algorithm}
\end{algorithm}

\textbf{Step~1: Constructing Domain Knowledge.}
A Knowledge Base (KB) includes a collection of concepts organized into categories, instances for every concept, and the relationship between these concepts~\cite{DataSynapse}.
We have constructed KBs consisting of different words related to radicalization criteria through which we can distinguish whether an individual is likely to be at risk of being radicalized in social networks or not. Here, we have tried to collect all criteria (\emph{illustrated as C}) which have been provided by domain experts~\cite{kruglanski2014psychology,li2011trust,bermingham2009combining,radical,bartlett2012edge,king2011radicalization,saif2017semantic} to assess the level of each person's radicalization. We consider thirteen criteria which can be assessed by using context-aware and activity-based approaches which were mentioned before. We describe all criteria in the following:

\emph{C1} Disappointment: this criterion tries to determine whether a person is disappointed or not by assessing the number of negative and swear words~\cite{lara2017measuring}.

\emph{C2} Introversion. This criterion can be measured by the length of the sentences and use of ellipses in the tweets ~\cite{lara2017measuring}.

\emph{C3} Discrimination. This factor can be measured by using keywords like sick, hate, detest, loathe, discrimination, people, racism, religion, etc~\cite{lara2017measuring}.

\emph{C4} Positive ideas about religion. This criterion can be assessed by using some keywords like religious, Muslim, Islam, Christian, good, best, nice, positive, excellent, happy, etc~\cite{lara2017measuring}.

\emph{C5} Showing negative thoughts about Western community. Radicalization indicates a movement in the direction of supporting or enacting radical behaviour~\cite{kruglanski2014psychology}. Some research has shown that those individuals in the social networks such as Twitter who write negative tweets against Western societies are more likely to support radical behavior. This perception can be stated by using negative thoughts about Western community like Western, people, hate, etc~\cite{lara2017measuring}.

\emph{C6} Mental Health. Some research findings have confirmed that there is a correlation between people who have anxiety and depression with those who have been radicalized~\cite{bhui2014might,bhui2014violent}.

\emph{C7} Racism: Those who are racist are more likely to be radicalized. Therefore, we consider different keywords regarding racism into radicalization issue~\cite{gilperez2017initial}.

\emph{C8} Political ideology. There are various kinds of political ideologies such as anarchism, absolutism, liberalism, conservatism, socialism and etc. Research findings have confirmed that radicalization issue makes individuals get interested in political matters.~\cite{borum2014psychological}.

\emph{C9} Educational level. Some researchers believe that radicalized people are usually well-educated~\cite{horgan2008profiles,monahan2012individual}. They also would argue that students are more likely to be radicalized than employees~\cite{bhui2014violent}.

\emph{C10} Origin. Those who become radicalized in the Western society are usually second~\cite{mastors2014omar}, or third generation foreigners~\cite{horgan2008profiles} in a Western society. However, there are some cases where first generation foreigners have got radicalized~\cite{bhui2014violent}.

\emph{C11} Personality traits. There are some personality features which are highly related to radicalization process such as being sensitive, having bipolar personality, adventure seeker (these people may take dangerous behavior for a sense of adventure~\cite{mastors2014omar,monahan2012individual}).

\emph{C12} Attitude towards terrorism. Some experts believe that violent and aggressive behavior is a common risk factor for being radicalized~\cite{borum2014psychological,doosje2016terrorism,pressman2016internet,trujillo2016psychometric}

\emph{C13} Psychological factors. There are some psychological variables which may lead an individual into radicalization process such as personal suffering (e.g. getting divorced, loss of a family member, sense of belonging), having low self-esteem, being isolated from the society, searching for purpose, feeling embarrassment and injustice~\cite{gilperez2017initial}.

Of all above criteria, our experiments have shown that some of the radicalization criteria have an important role in affecting influential users to be radicalized. Hence, we propose a formula based on our experiments to compute the average rate of radicalization for each influential user. The formula is proposed in the following:

\begin{equation}
Radicalization = Average \{C_2, C_3, C_4, C_5, C_8, C_{11}, C_{12}\}
\end{equation}
Where
\begin{equation}
C_j = (\frac{\sum_{i=1}^{n}C_j(for seed size = i)}{n})
\end{equation}

and $n$ is the maximum number of seed size which is used in PSO algorithm.
Figure~\ref{fig:Demo} illustrates a software prototype, implemented to analyze the users' behavioral disorder and to identify radicalized people on online social networks.

\textbf{Step~2: Context Analytics.}
In this part, the results of algorithm~1 which were the influential users in a social network (seeds) are given to algorithm~2 as inputs, then we give different knowledge bases (which contain dictionaries consisting of different radicalization criteria that show a person is likely to radicalize others) to algorithm 2 as inputs as well. The knowledge bases are explained in the following. Then we use a function called $getwords()$ to read all the words from each knowledge base, then we use another function called $getrate()$ to calculate the rate of each knowledge base from each tweet of influential users. The rate of each knowledge base means that how many words in each knowledge base exist in the tweets of each user.
The schema of the second algorithm is shown in the following as algorithm 5.
Figure~~\ref{fig:Figure3} illustrates a typical scenario for this algorithm.

\chapter{Experiments and Evaluation}
\label{chap:experiments}

In this chapter, we present the dimensions for evaluation which will be used to evaluate the performance of our proposed algorithms.
To demonstrate the applicability of these dimensions, we provide the evaluation and discussion of our proposed algorithm's results and compare the results with Memetic algorithm because Memetic algorithm has shown the best result regarding effectiveness and efficiency so far~\cite{7515281}.
We have also discovered a set of \emph{benchmarking datasets} which can help readers in the identification of significant and outstanding issues for further research.


\section{Benchmarking Datasets}
\label{Datset}

Measuring the effectiveness of our proposed algorithms is difficult and requires large datasets providing adequate level of ambiguity (the ability to represent multiple interpretations) and precise ground-truth (the accuracy of the training set's classification for supervised learning techniques).
To accomplish this task, we use a large Twitter dataset (around 15 million Tweets) and a labeled Twitter dataset contains a set of radicalized people. We also use the well known Dolphin dataset to evaluate the effectiveness of the proposed Influence Maximization algorithm.
Details of these datasets has been illustrated in Table~4.1


\begin{table}[ht]
\caption{Datasets and their description}\label{tab1}
\begin{tabular}{|l|l|}
\hline
Dataset &  Description\\
\hline
\textbf{Twitter Dataset}~\cite{beheshti2018datasynapse}         &  The 3-month tweets in Australia which consist of 14,976,862 \\
                    & tweets. The Twitter dataset is modeled as a graph of nodes \\
                    & (users/actors in twitter) and the relationships among them.\\
\textbf{Labeled Twitter Dataset} & Provided by Kaggle~\cite{Kaggle}, the dataset consists of 17000 tweets\\
                                 & from a number of Twitter accounts of ISIS proponents related to \\
                                 & the Paris invasion in 2015.\\
\textbf{Dolphin}~\cite{lusseau2003bottlenose} &   Is a small social network which describes the relationship between\\
        &  62 dolphins living in New Zealand. The nodes of the dataset indicate\\
        &  dolphins and edges  represent relationship among these dolphins.\\
\hline
\end{tabular}
\end{table}

\setlength{\arrayrulewidth}{0.4mm}
\setlength{\tabcolsep}{8pt}
\renewcommand{\arraystretch}{1.2}
\begin{table}
\centering
\begin{tabular}{|l|c|c|}
\hline
\textbf{Parameter} & \textbf{Meaning} & \textbf{Value}  \\
\hline
$Cog_{eff}$ & Cognitive coefficient term in Algorithm(1) & 1.5 \\
\hline
$Soc_{coe}$ & Social coefficient in Algorithm(1) & 100 \\
\hline
$\theta$ & The constant term in (3.3) & 4 \\
\hline
$\lambda$ & The magnification term in (3.3) & 10 \\
\hline
$sim_{thresh}$ & A threshold term in Algorithm(3) & 0.6 \\
\hline
maxgen & Maximum generation & 50 \\
\hline
Population size & population size & 20 \\
\hline
\end{tabular}
\caption{\label{tab:detail1}provides the parameters in first algorithm.}
\end{table}

\section{Experiment}
\label{Comparing Algorithms}

We did the comparison of our first algorithm with Memetic algorithm~\cite{7515281} using Twitter dataset to evaluate performance of our algorithm. All tests are done using the IC model. To compare the performance of our algorithm, we calculate the fitness function for different number of generations and different number of seeds (influential nodes) in our algorithm and compare the results with memetic algorithm. All the experiments are run on a machine with 3.40 GHz Inter Core i7 and 8.00 GB RAM. The experimental factors of both algorithms are given in Table 4.2.

\begin{figure}[ht]
  \begin{minipage}[b]{0.5\linewidth}
    \includegraphics[width=1\linewidth]{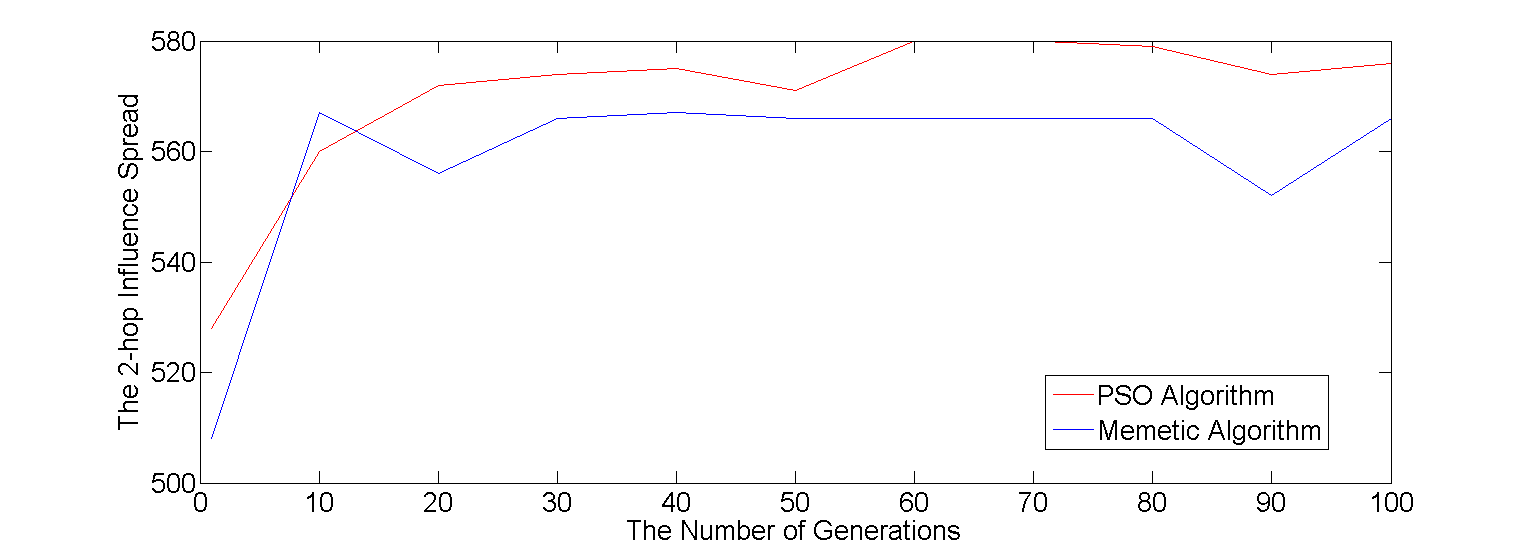}
    \caption{Comparsion between PSO
    and Memetic using the Dolphin dataset}
  \end{minipage}
  \begin{minipage}[b]{0.5\linewidth}
    \includegraphics[width=1\linewidth]{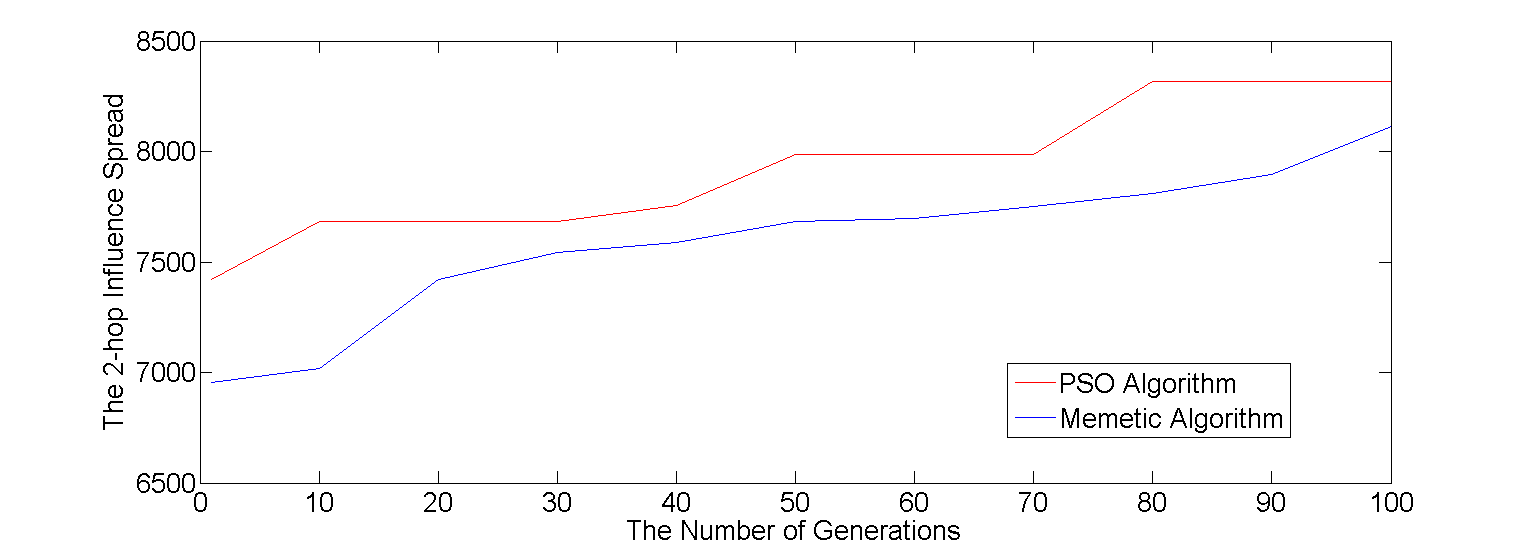}
    \caption{Comparsion between PSO and Memetic using the Twitter dataset }
  \end{minipage}
  \begin{minipage}[b]{0.5\linewidth}
    \includegraphics[width=1\linewidth]{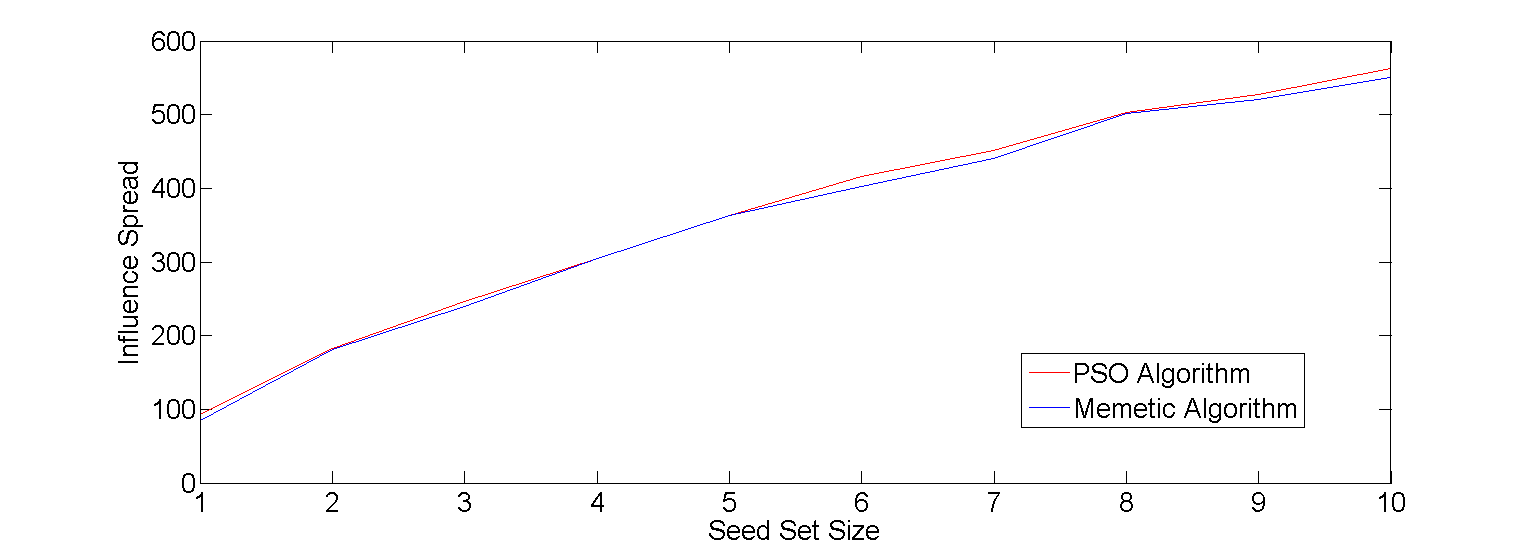}
    \caption{Influence spread of PSO and Memetic using the Dolphin dataset}
  \end{minipage}
  \hfill
  \begin{minipage}[b]{0.5\linewidth}
    \includegraphics[width=1\linewidth]{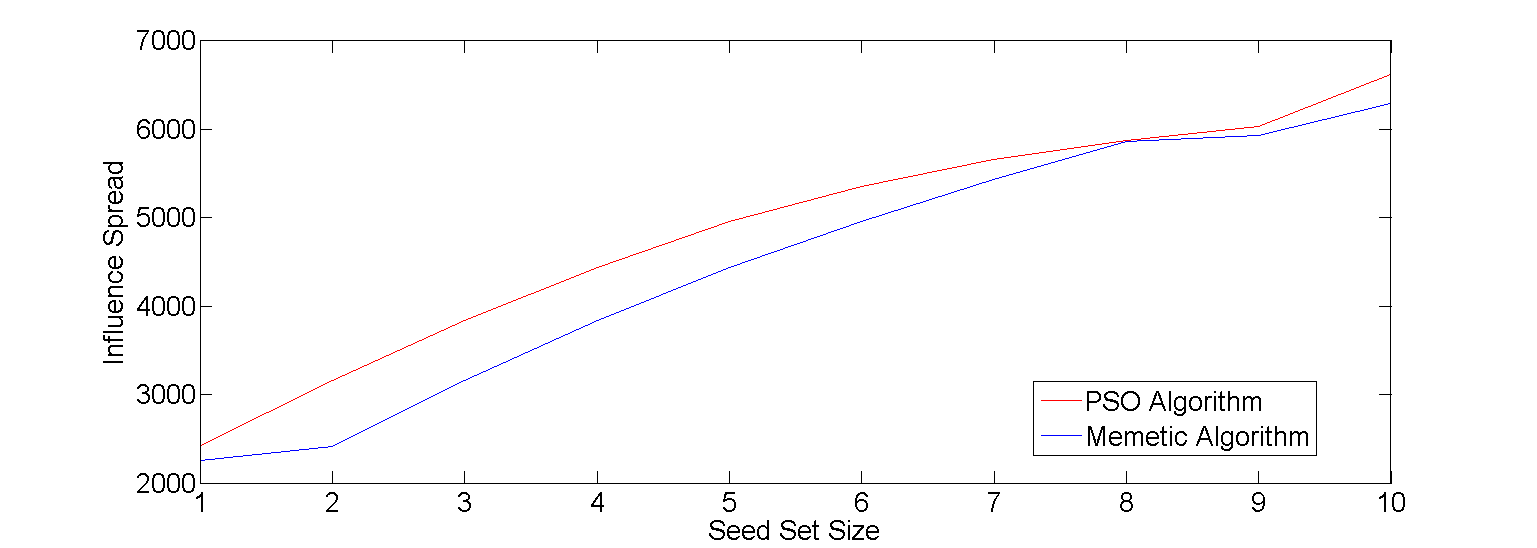}
    \caption{Influence spread of PSO and Memetic using the Twitter dataset}
  \end{minipage}
\end{figure}

Firstly, we test PSO and Memetic algorithm on two real-world dataset, the Dolphin network and the Twitter dataset. In these experiments, we run both algorithms by different number of generations. The results are shown in Figs.~4.1 and~4.2, where the red and blue lines illustrate the 2-hop influence spread obtained by PSO and Memetic algorithms respectively with generations ranging from 1 to 100 on the Dolphin and Twitter dataset. Both PSO and Memetic use local search algorithm so as to find the optimal solution.
The results show that local search can outperform the quality of solution and increase the convergence.

Fig.~4.3 shows the influence spread of two algorithms on the Dolphin Dataset using the IC model in which x-axis indicates the seed set size and y-axis signifies the influence spread by performing 50 generations. As can be seen from Fig.~4.3, the contrast in the influence spread of PSO and Memetic algorithm is negligible although PSO shows the better result. Therefore, in small social networks like the Dolphin, both PSO and Memetic can perform similarly with different number of seed size.
Fig.~4.4 demonstrates the influence spread of two algorithms on the Twitter Dataset using the IC model in which x-axis shows the seed set size and y-axis indicates the influence spread by performing 50 generations. As obvious from Fig.~4.4, the influence spread of the two algorithms is very different. The result of PSO is quite better than that of Memetic algorithm. Thus, the bigger size of a social network (Twitter social network), the better result PSO algorithm can obtain compared with Memetic algorithm.

As summerized from the experimental results, the PSO algorithm is a decisive factor in increasing the convergence and finding the best solutions though it can obtain a better result when we have a bigger network with more nodes. Besides, PSO algorithm with local search procedure can resolve the influence maximization problem efficiently on social networks with various sizes.
After identifying the influential users, we use the second algorithm to obtain those influential users (seeds) who have the ability to make other users radicalized. In this part, we used the Twitter dataset (3-month tweets in Australia) and a labeled Twitter dataset (provided by Kaggle~\cite{Kaggle}) including 17000 tweets related to ISIS proponents. We also use different knowledge bases related to radicalization factors to find that which seeds from the first algorithm is able to radicalize the other users in both datasets.

Our second algorithm 
shows the following result.
With first Twitter dataset consisting of different number of tweets (i.e, 10,000 tweets, 100,000 tweets, 1,000,000 tweets and 10,000,000 tweets), the experimental results are listed in Table 4.3.
With second Twitter dataset consisting of 17000 tweets of ISIS proponents, the radicalization factors are computed based on different seed size in Table 4.4. Based on formula~3.12 for computing the radicalization rate, we calculated the average radicalization rate for Twitter dataset provided by Kaggle.
Based on the experiments in table~4.4, we calculate the average radicalization rate for each influential user as follows.\\
$C_2= (10+10+10)/5 =6$ , $C_3=14.3/5 = 2.86$ , $C_4= (0.5 + 27 + 11+ 5 + 11)/5 = 10.9$\\
$C_5= (30.7 + 30.7 + 30.8 + 15.4 + 15.4)/5 =24.6 $ , $C_8= 30/5 = 6$ , $C_{11}= 28.6/5 = 5.72 $\\
$C_{12}= 3/5 = 0.6 $ , $Radicalization = Average\{6, 2.86, 10.9, 24.6, 6, 5.72, 0.6\} = 8.097\% $

\setlength{\arrayrulewidth}{0.4mm}
\setlength{\tabcolsep}{3.2pt}
\renewcommand{\arraystretch}{.7}

\begin{table}[]
\caption{\label{tab:detail2}provides the results from the Twitter dataset consisting of 3-month tweets in Australia.}
\begin{tabular}{|c|l|l|l|l|l|}
\hline
\multicolumn{6}{|c|}{\textbf{Seed Size}}                                                                                          \\ \hline
\multirow{13}{*}{\textbf{Radicalization factors}} &                                  & 10,000 & 100,000 & 1,000,000 & 100,000,000 \\ \cline{2-6}
                                                  & Introversion                     & 0      & 0       & 0         & 0           \\ \cline{2-6}
                                                  & Discrimination                   & 0      & 0       & 0         & 0           \\ \cline{2-6}
                                                  & Pos. Ideas about religion        & 0      & 0       & 0         & 0           \\ \cline{2-6}
                                                  & Neg. ideas about Western society & 0      & 0       & 0         & 0           \\ \cline{2-6}
                                                  & Mental health                    & 0      & 0       & 0         & 0           \\ \cline{2-6}
                                                  & Racism                           & 0      & 0       & 0         & 0           \\ \cline{2-6}
                                                  & Political ideology               & 0      & 0       & 0         & 0           \\ \cline{2-6}
                                                  & Educational level                & 0      & 0       & 0         & 0           \\ \cline{2-6}
                                                  & Origin                           & 0      & 0       & 0         & 0           \\ \cline{2-6}
                                                  & Personal traits                  & 0      & 0       & 0         & 0           \\ \cline{2-6}
                                                  & Terrorism                        & 0      & 0       & 0         & 0           \\ \cline{2-6}
                                                  & Psychological factors            & 0      & 0       & 0         & 0           \\ \hline
\end{tabular}
\end{table}


\setlength{\arrayrulewidth}{0.4mm}
\setlength{\tabcolsep}{3.2pt}
\renewcommand{\arraystretch}{.7}
\begin{table}[]
\caption{\label{tab:detail3} provides the results from Twitter dataset consisting 17000 tweets of ISIS proponents with different seed size.}
\begin{tabular}{|c|l|l|l|l|l|l|}
\hline
\multicolumn{7}{|c|}{\textbf{Seed Size}}                                                                                               \\ \hline
\multirow{13}{*}{\textbf{Radicalization factors}} &                                  & 1       & 2       & 3       & 4       & 5       \\ \cline{2-7}
                                                  & Introversion                     & 10\%    & 10\%    & 10\%    & 0\%     & 0\%     \\ \cline{2-7}
                                                  & Discrimination                   & 0       & 0       & 0       & 0       & 14.30\% \\ \cline{2-7}
                                                  & Pos. Ideas about religion        & 0.50\%  & 27\%    & 11\%    & 5\%     & 11\%    \\ \cline{2-7}
                                                  & Neg. ideas about Western society & 30.70\% & 30.70\% & 30.80\% & 15.40\% & 15.40\% \\ \cline{2-7}
                                                  & Mental health                    & 0       & 0       & 0       & 0       & 0       \\ \cline{2-7}
                                                  & Racism                           & 0       & 0       & 0       & 0       & 0       \\ \cline{2-7}
                                                  & Political ideology               & 10\%    & 0       & 0       & 10\%    & 10\%    \\ \cline{2-7}
                                                  & Educational level                & 0       & 0       & 0       & 0       & 0       \\ \cline{2-7}
                                                  & Origin                           & 0       & 0       & 0       & 0       & 0       \\ \cline{2-7}
                                                  & Personal traits                  & 0       & 28.60\% & 0       & 0       & 0       \\ \cline{2-7}
                                                  & Terrorism                        & 0       & 3\%     & 0       & 0       & 0       \\ \cline{2-7}
                                                  & Psychological factors            & 0       & 0       & 0       & 0       & 0       \\ \hline
\end{tabular}
\end{table}

\chapter{Conclusion and Future directions}
\label{chap:conclusion}

In this chapter, we conclude the contributions of this dissertation and discuss some
future research directions to build on this work.

\section{Concluding Remarks}

Actions of violent extremists menace humanities' principles and core values which include the rule of law and human rights.
While radical thinking itself is not problematic, it poses a threat to national security when it leads to violence.
Major barriers to the effective understanding of extremist and criminal behaviors on social networks is the ability to understand the context as well as eventual influence that an influence campaign may have, given an understanding of the mode of information dissemination and indicators that identify the use of artificial promotion of a theme.
To overcome this challenge, we proposed a social data analytics pipeline, namely iRadical, to enable analysts engage with social data to explore the potential for online radicalization.
We introduced the novel concept of Social Personality Graph, to model and analyze factors that are driving extremist and criminal behavior.
We presented algorithms to analyze the user activity patterns to learn how influence flows in social networks considerying the features extracted into the personality graph.
We presented the evaluation results on the performance and the quality of the results using real-world and synthetic data.

\section{Future Directions}

In this dissertation, we have investigated the problem of finding a small subset of nodes in a social network which can maximize the propagation of influence as well as the discovery, interpretation, and communication of meaningful patterns in social data to enable analysts to understand patterns of behavioral disorders.
We believe that this is an interesting area of research and has the potential of growing in the coming years. 
We summarize 
some research directions in this area as follows.

\textbf{Social Data Analytics.}
Social data analytics have become an integral part for governments and organizations.
For instance, for the past few years, governments have used data analytics to extract useful information from the open data and have used it to provide tailored advertisements during elections and for improving different services for the people. It has also been used for predicting intelligence activities and for improving public health and national security~\cite{DataSynapse,DBLP:conf/adc/MaamarSBB15,DBLP:conf/icsoc/SunBBB15,DBLP:conf/bpm/BeheshtiBNS11}.

\textbf{Social Data Curation.}
A major challenge in analyzing social data is the conversion of the raw data generated by social actors into curated data~\cite{beheshti2017automating}, i.e. contextualized data and knowledge which is maintained and made available for use by applications and end-users.
Novel techniques to extract, enrich, annotate and summarize domain specific data and knowledge would be a vital asset to analyze patterns of behavioral disorder.

\textbf{Influence Maximization.}
Finding a few number of nodes in a social network which could increase the propagation of influence, would be vital for understanding and analyzing the influence dimension in criminal and extremist activities. We plan to link trust prediction~\cite{DBLP:conf/wise/GhafariYBO18,DBLP:conf/wise/GhafariYBO18a,DBLP:conf/wise/YakhchiGB18} approaches to the influence maximization to understand and analyze the influence dimension.



\backmatter


\bibliography{references}

\end{document}